\documentclass[12pt]{article}
\usepackage{pdflscape}
\usepackage{amsmath,multirow,amsfonts,amssymb,bm}
\usepackage[T1]{fontenc}
\usepackage[cp1250]{inputenc}
\usepackage{authblk}
\usepackage{graphicx}
\usepackage[table]{xcolor}

\numberwithin{equation}{section}

\newcommand{\el}{\ \\\nonumber}
                                        
\newcommand{\sh}[1]{#1\hskip-7pt \diagup}

\newcommand{\bracket}[3]{\left<#1\left|#2\right|#3\right>}

\newcommand{\ups}[2]{u_{#2}(\bm{#1})}

\newcommand{\oups}[2]{\overline{u}_{#2}(\bm{#1})}

\newcommand{\mintt}[1]{\int\hspace{-5pt}\frac{d^3p_{#1}}{(2\pi)^3}}
\newcommand{\dpt}{(2\pi)^3}
\newcommand{\dpc}{(2\pi)^4}
\newcommand{\tr}{{\mathrm{Tr}}}
\newcommand{\opsi}{\overline{\psi}}

\newcommand{\gfiv}{\gamma^5}
\newcommand{\gmu}{\gamma^\mu}

\newcommand{\gal}{\gamma^\alpha}
\newcommand{\gbe}{\gamma^\beta}

\newcommand{\Dmu}{\partial^\mu}

 

\title{On Impact of Nuclear Effects on Weak Pion Production in Sub 1~GeV Energy Region}

\author[1]{Jan T. Sobczyk\footnote{On leave from the Institute of Theoretical Physics, Wroc\l aw University }}
\author[2]{Jakub \.{Z}muda}
\affil[1]{\it Fermi National Accelerator Laboratory, Batavia, Illinois 60510, USA.}
\affil[2]{\it Institute of Theoretical Physics, Wroc\l aw University, pl. M. Borna 9, 50-204, Wroc\l aw, Poland}

\date{}

\begin{document}
\maketitle

\section{Abstract}
%An impact of nuclear effects on charged-current single pion production scattering off $^{12}C$ is investigated 
%for neutrino energies up to $1$~GeV. In-medium modifications of the $\Delta(1232)$ resonance properties
%as well as an effective field theory nonresonant background contribution are discussed. A model of Nieves et al 
%\cite{Nieves:2011pp} is further developed by performing exact integration avoiding several 
%approximations.

Charged-current single pion production in scattering off $^{12}C$ is investigated for neutrino energies up to $1$~GeV.
An impact of nuclear effects with in-medium modifications of the $\Delta(1232)$ resonance properties as well as an effective field theory
nonresonant background contribution are discussed. Dependence of the fraction of $\Delta(1232)$ decays into $npnh$ states  on
incident neutrino energy is estimated. A model of Nieves {\it et al.} \cite{Nieves:2011pp} is further developed by performing exact integration
avoiding several approximations. The effect of exact integration is investigated both for double-differential and total
neutrino-nucleus cross sections.

%%%%%%%%%%%%%%%%%%%%%%%%%%%%%%%%%%%%%%%%%%%%
\section{\label{sec:level1}Introduction}
%%%%%%%%%%%%%%%%%%%%%%%%%%%%%%%%%%%%%%%%%%%
%
There has been a lot of effort to understand better the single pion production (SPP) reactions in neutrino-nucleon 
and neutrino-nucleus scattering. 
Motivations for these studies come from neutrino oscillation experiments and their demand to reduce systematic errors.
In a few GeV energy region characteristic for experiments like T2K, MINOS, NOvA, MiniBooNE and MicroBooNE 
the SPP channels account for a large fraction of the cross section (at $1$~GeV on an isoscalar target $\sim 36$\%). 

In the neutrino experiments one typically measures charged current quasi-elastic (CCQE) events. 
The signal events have no pions in the final state. In the case of SPP 
reaction on nuclei there is a significant probability that a pion produced on a bound nucleon
gets absorbed, and such events contribute to the CCQE 
background. The MiniBooNE experiment has put a lot of effort to develop  
methods to estimate and subtract that background from the CCQE data sample \cite{AguilarArevalo:2010zc}. 
A data/MC correction function was introduced based on the investigation of events with a single pion 
in a final state. However,
it is not clear how legitimate was to assume that the same function can be applied to hypothetical
pion absorption events. A necessity to use the large correction function can be an indication
that the pion production model implemented in NUANCE Monte Carlo event generator suffers from deficiencies \cite{problems_RS}. 
Another well known instance of relevance of pion production channels is 
neutral current $\pi^0$ production. Neutral pions give rise to events which mimic
$\nu_\mu\rightarrow \nu_e$ signal (it happens if one of the two photons from the $\pi^0$ decay remains unidentified). 
In the nuclear medium intermediate virtual resonance states, leading normally 
to single pion production, may get absorbed by nucleons. The pionless $\Delta$ decays events
contribute to multinucleon ejection final states and can be confused with the genuine CCQE events because the 
knocked out
protons are usually not analyzed at all. The multinucleon knock-out contamination has an impact on the 
neutrino energy reconstruction and should be accounted for in the 
neutrino oscillation experimental analysis \cite{en_rec}.
Weak single pion production processes are also important for the hadronic physics. They provide a
valuable information both on the dynamical structure of the nucleon resonances and nonresonant contributions.
This information is complementary to what is known from the electro- and photoproduction studies.
This topic was studied in the MAINZ, BONN and TJNAF 
laboratories. The results give a good insight on the electromagnetic structure of nucleon resonances production,
see for example the latest paper on the subject \cite{Tiator:2011pw} and the underlying unitary isobar model 
for pion electroproduction \cite{Drechsel:2007if}. The above mentioned analysis includes also a variety of low 
lying resonances beyond $\Delta(1232)$. Three of them: $P_{11}(1440)$, $D_{13}(1525)$ and $S_{11}(1535)$ 
may be relevant for the understanding of pion production process in the neutrino experiments, like T2K or NOvA. Electro-
and photoproduction experiments can also serve as a valuable source of information on final state interactions (FSI) 
effects which are
universal for all pion production experiments though a number of such studies is limited.

Recent experimental results on the charge current SPP reactions 
come mainly from the K2K (\cite{Rodriguez:2008aa}, 
\cite{Mariani:2010ez}) and MiniBooNE experiments (\cite{AguilarArevalo:2010bm}, \cite{AguilarArevalo:2010xt}). 
Unfortunately, the analysis of the underlying fundamental physical processes of  pion production on nucleons is 
obscured by nuclear effects. There is an important impact of the nuclear medium on a primary interaction as well as 
on a redistribution of physical channels by FSI inside the
target nucleus. The nuclear physics uncertainties are so large that MiniBooNE did not attempt to
measure the characteristics of neutrino-{\it nucleon} SPP process and published the cross sections results with
all the nuclear effects included (the signal events are those with a single pion leaving a nucleus). 
A more valid information about the nucleon-$\Delta$ transition form factors and the  effects 
of the nonresonant background can be extracted from the data obtained on light targets like hydrogen or deuterium. 
The results from two old low statistics dedicated bubble chamber experiments, 
ANL (\cite{Barish:1978pj}, \cite{Radecky:1981fn}) and 
BNL (\cite{Kitagaki:1990vs}), are still used in the verification of dynamical 
pion production models and weak $N-\Delta$ transition form-factors (see for example \cite{Hernandez:2007qq}, 
\cite{Lalakulich:2010ss}).

In the theoretical discussions of weak SPP two main concerns are: 
description of the non-resonant background and handling of the nuclear effects.
For the neutrino energies of about $1$~GeV a dominant pion production channel is that via $\Delta$ excitation.
However, as clearly follows from the ANL and BNL experiments, there is an important
non-resonant contribution that cannot be neglected though much harder to include in the computations. The nuclear
effects include the $\Delta$ in-medium self energy which leads to a substantial fraction of pionless decays. 
An impact of the nuclear medium on the vector and axial $\Delta$ excitation form-factors is unknown and currently
there is no method of measuring it. 
The problem of charged-current SPP on nuclei assuming $\Delta$ dominance model with
many-body effects from \cite{Oset:1987re} has been addressed in \cite{Singhy}. The 
computations have shown a significant reduction of the pion production cross-section due 
to the in-medium effects. The fraction of $\Delta$ pionless decays has a rather mild dependence
on the incident neutrino energy. All these calculations did not include any kind of nonresonant background. 
An assumption of the constant fraction of pionless $\Delta$ decay is implemented in neutrino Monte Carlo
event generators (NUANCE, NEUT) and the nonresonant dynamics is modelled in a simplified manner within an old
fashioned Rein-Sehgal model which is known to suffer from other deficiences as well \cite{problems_RS}.
In this paper the full model of weak SPP on nuclei based on \cite{Hernandez:2007qq} is used for the neutrino-nucleus
scattering following the approach of  \cite{Nieves:2011pp}. The impact of the nonresonant background is discussed.
A lower bound for $np-nh$ contribution coming from the pionless $\Delta$ decays is estimated. 
The goal of this paper is to present the predictions from the sophisticated theoretical model for 
SPP in such away that they
can be used in the evaluation of the systematic errors by experimental groups. 
An impact of various ingredients of the model
on the final results will be presented as well.

Our model is based on the papers by Nieves et al.(\cite{Hernandez:2007qq}, \cite{Nieves:2011pp})
and we aim to further develop their approach. Our most important contribution is a prescription
how to perform many integrals in an exact way. Thanks to that we avoid  not easy to control
approximations. It turns out that the approximations used in \cite{Nieves:2011pp} 
do not work well in the case of double-differential cross sections, while
for total cross sections they produce results close to the exact ones.
We have found also, that the assumption of a constant fraction of pionless $\Delta$ decay cannot be applied for experiments with
large flux contribution from $E_\nu<1\ \mathrm{[GeV]}$ and that the ratio of muon to electron
(anti-)neutrino total crosss section does not depend on the medium modifications of $\Delta(1232)$ resonance.

The paper is organized as follows: in Section \ref{sec:genform} we discuss the general formalism of SPP on atomic
nuclei. The dynamical model of SPP is reviewed in Section \ref{sec:sppdyn} and the nuclear medium effects
are discussed in Section \ref{sec:mediumdel}. In Section \ref{sec:numer} we briefly introduce the numerical procedures
and in Section \ref{sec:res} we present our main results that are then discussed in Section \ref{sec:discuss}

\section{Theoretical Description of Pion Neutrinoproduction on Atomic Nuclei}
\label{sec:genform}
The theoretical approach presented in this paper is based on the general scheme described in \cite{Nieves:2011pp}. 
The basic cross-section formula for the electromagnetic or weak charged-current lepton inclusive differential cross section is:
\begin{eqnarray}\label{eq:xs1}
		\frac{d^3\sigma}{d\Omega'dE'}&=&F_l(Q^2)\frac{|l'|}{|l|}\int\vspace{-2pt}d^3r 
L_{\mu\nu} W^{\mu\nu}(\rho(\vec r) )\el
&&\el
		F_l(Q^2)&=&\left\{\begin{array}{c} \frac{2\alpha^2}{Q^4},\ \mathrm{electrons}\\ 
\frac{G_F^2\cos^2\theta_C}{4\pi^2},\ \mathrm{(anti)neutrinos} \end{array}\right.\el 
&&\el
		L_{\mu\nu}&=&\left\{\begin{array}{c} l_\mu l'_\nu +l'_\mu l_\nu -g_{\mu\nu} ll',\ 
\mathrm{electrons}\\ l_\mu l'_\nu +l'_\mu l_\nu -g_{\mu\nu}ll'\pm i \epsilon_{\mu\nu\alpha\beta}l'^\alpha l^\beta,\ \mathrm{(anti)neutrinos} \end{array}\right.
\end{eqnarray}
For the weak interactions the Fermi contact is $G_F=1.1664*10^{-11}/MeV^2$
and the cosine of Cabbibo angle is $\cos(\Theta_C)=0.974$. Furthermore $l^\mu$ and $l'^\mu$ denote initial/final
lepton four-momenta, $q^2=-Q^2=(l-l')^2$ is the squared four-momentum 
transfer. In the laboratory frame we assume the momentum transfer to be directed along the Z-axis 
and the scatttering to take place in the X-Z plane. The local density approximation is adopted with $\rho(\vec r)$
being the nuclear matter density. The parameterization we adopted in the
numerical computations as well as several other technical details is given in
Appendix \ref{sec:ldapar}. 

The cross section can be re-expressed in terms of the gauge boson self-energy in nuclear medium, 
it is readily done by a substitution:
\begin{eqnarray}\label{eq:xs2}
L_{\mu\nu} W^{\mu\nu}(\rho(\vec r) )=-\frac{1}{\pi}\Im\left[L_{\mu\nu}\Pi^{\mu\nu}(q,\rho(\vec r))\right].
\end{eqnarray}
The polarization tensor $\Pi^{\mu\nu}$ has a dimension of (energy)$^3$. 
After multiplying it by an appropriate external couplings and performing the spatial $d^3r$ integration 
one gets a representation of the gauge boson self-energy. 
It can be evaluated by adding contributions from Feynman diagrams representing 
various processes, with nucleon loops having
momentum cutoffs given by local Fermi momentum

A dominant SPP part is in the many body language denoted as $1p1h1\pi$ (contributions from $2p2h1\pi$ 
and more complicated final states is assumed to be small): there is one pion and one nucleon-hole pair ($1p1h$) 
in the final state.
The corresponding contribution to polarization tensor can be represented as:
\begin{eqnarray}
\label{eq:pimunu1p1h1pi}
-i\Pi^{\mu\nu}_{1p1h1\pi}&=&\sum_{iso}\hspace{-2pt}\mint{}\hspace{-2pt}\int\hspace{-5pt}\frac{d^4k}{\dpc}iD_{\pi}(k)iG_N(p)iG_{N'}(p\hspace{-2pt} + \hspace{-2pt} q \hspace{-2pt} -\hspace{-2pt} k)\tr\hspace{-2pt}\left[\hspace{-2pt} A^{\mu\nu}_{1p1h1\pi}(p,q,k)\hspace{-2pt} \right].
\end{eqnarray}
The hadronic tensor $A^{\mu\nu}_{1p1h1\pi}$ is defined as:
\begin{eqnarray}
A^{\mu\nu}_{1p1h1\pi}&=&\sum_{s,s'}\bracket{N'(p',s')\pi(k)}{j^\mu_{cc}}{N(p,s)}(\bracket{N'(p',s')\pi(k)}{j^\nu_{cc}}{N(p,s)})^\dag
\end{eqnarray}
In (\ref{eq:pimunu1p1h1pi}) $G_N$ denotes the nucleon propagator:
\begin{eqnarray}
\label{eq:nucprop}
G_N(p)=\frac{1}{p^0+E(p)+i\epsilon}\left( \frac{n_N(\vec{p})}{p^0-E(p)-i\epsilon}+\frac{1-n_N(\vec{p})}{p^0-E(p)+i\epsilon}\right)
\end{eqnarray}
with $n_N(\vec{p})$ being the occupation numbers for nucleon of isospin $N$. 
In the Fermi gas (FG) model $n_N(\vec{p})$ is a Heaviside step function $\Theta(|\vec{p}|-k_F^N)$. 
The $\bracket{N'(p',s')\pi(k)}{j^\mu_{cc}}{N(p,s)}$ are transition amplitudes between initial nucleon state
with spin $s$ and four-momentum $p$ and final state containing pion with four-momentum $k$ and nucleon 
with four-momentum $p=p+q-k$ and spin $s'$. After inserting the nucleon propagators into polarization tensor 
we obtain the following expression:
\begin{eqnarray}
\label{eq:intform}
-\frac{1}{\pi}\Im\left(\Pi^{\mu\nu}_{1p1h1\pi}L_{\mu\nu}\right)&=&\sum_{iso}\hspace{-2pt}\mintt{}\hspace{-2pt}
\int\hspace{-5pt}\frac{d^3k}{\dpt}
\frac{1}{8E_\pi(k) E(p) E(p')}\el 
&&\el
& &\left[ n_N(p)(1\hspace{-2pt} -
\hspace{-2pt} n_{N'}(p'))+ n_{N'}(p')
(1\hspace{-2pt} -\hspace{-2pt} n_{N}(p))\right]\el 
&&\el
& &\delta(E(p')- q^0 
\hspace{-2pt} +\hspace{-2pt} E_\pi\hspace{-2pt} -\hspace{-2pt} E(p))\tr\hspace{-2pt}\left[\hspace{-2pt} A^{\mu\nu}_{1p1h1\pi}(p,q,k)\hspace{-2pt} \right]L_{\mu\nu}
\end{eqnarray}
with the nucleon energy $E(p)=\sqrt{\vec{p}^2+M^2}$ and the final pion energy $E_\pi(k)=\sqrt{\vec{k}^2+m_\pi^2}$. 
Taking into account, that the pion may carry a charge and the nucleus atomic number can
be changed, one can establish the threshold corrected energy transfer (as for the quasielastic
peak):
\begin{eqnarray}
\label{eq:tq0}
\tilde{q}^0=q^0-Q_{corr}+\Delta_{E_F},\quad \Delta_{E_F}\equiv E_F^{N}-E_F^{N'}.
\end{eqnarray} 
In this way one accounts for the difference of rest masses of 
isotopes by subtracting the rest mass difference $Q_{corr.}$ and
different Fermi levels of protons and neutrons. 
We substitute $q^0\to\tilde{q}^0$ everywhere in the hadronic part of the polarisation tensor. 
An alternative approach for nuclear the binding energy is used by \cite{Nieves:2011pp} and
shortly explained in Appendix \ref{sec:ldapar}. 
In the isospin symmetric nuclei, like $^{12}C$
the exchange part of cross section given by the terms with $n_{N'}(p')(1-n_N(p))$
is negligibly small and thus we neglect it.

\subsection{Dynamics of Single Pion Production}
\label{sec:sppdyn}
\begin{figure}[htb]
\centering\includegraphics[height=4.2cm]{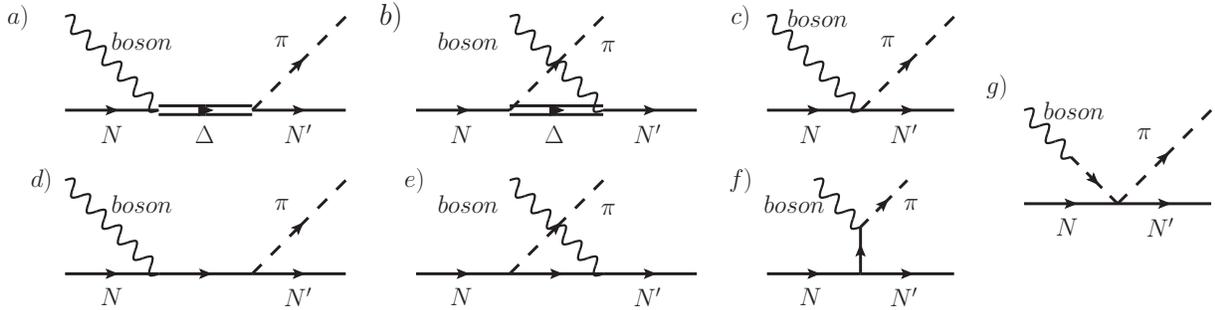}
\caption{Basic pion production diagrams from \cite{Hernandez:2007qq}: 
a) Delta pole ($\Delta$P), b) crossed  Delta pole (C$\Delta$P), c) contact term (CT), d) nucleon pole (NP), 
e) crossed nucleon pole (CNP), f) pion-in-flight (PIF), g) pion pole (PP) }\label{fig:mec}
\end{figure}

The dynamics is defined by a set of Feynman diagrams (Fig. \ref{fig:mec}) 
with vertices determined by effective chiral field theory \cite{Hernandez:2007qq}. 
The same set of diagrams describes also pion electroproduction, with the exception of pion pole (PP) diagram, 
which is purely axial.

After performing summations over nucleon spins we can rewrite the hadronic tensor as:
\begin{eqnarray}
A^{\mu\nu}&=&\tr\left[(\sh{p}'+M)s^\mu (\sh{p}+M)\gamma^0s^{\nu\dag}\gamma^0\right].
\end{eqnarray}
Reduced current matrix elements $s^\mu$ correspond to  weak transition amplitudes:
\begin{eqnarray}
\label{eq:smu}
\bracket{N'(p',s')\pi(k)}{j^\mu_{cc}}{N(p,s)}=\oups{p'}{s'}s^\mu\ups{p}{s}.
\end{eqnarray}
They are calculated to be (see: \cite{Hernandez:2007qq}):
\begin{eqnarray}
\label{eq:dp}
s^\mu_{\Delta P}&=&i C_{\Delta P} \frac{f^\ast}{m_\pi} \cos \Theta_C\frac{k^\alpha P_{\alpha\beta}(p+q)\Gamma^{\beta\mu}(p,q) }{(p+q)^2-M_\Delta^2 + iM_\Delta \Gamma((p+q)^2)} \\
\label{eq:cdp}
s^\mu_{C \Delta P} &=& i C_{C\Delta P} \frac{f^\ast}{m_\pi} \cos \Theta_C\frac{ \gamma^{0}\left[\Gamma^{\alpha\mu}(p-k,-q)\right]^\dag \gamma^0 P_{\alpha\beta}(p-k) k^\beta  }{(p-k)^2-M_\Delta^2 + iM_\Delta \Gamma((p-k)^2)}\\
\label{eq:np}
s^\mu_{NP} &=& -i C_{N P} \frac{g_A}{\sqrt{2} f_\pi} \cos \Theta_C\frac{ \sh{k}\gamma^5(\sh{p}+\sh{q}+M)  }{(p+q)^2-M^2 + i\epsilon}j^\mu_{CCN}(q)F_{\pi}(k-q)\\
\label{eq:cnp}
s^\mu_{CNP} &=& -i C_{CNP} \frac{g_A}{\sqrt{2} f_\pi} \cos \Theta_C j^\mu_{CCN}(q) \frac{ (\sh{p}-\sh{k}+M)\sh{k}\gamma^5  }{(p-k)^2-M^2 + i\epsilon}F_{\pi}(k-q) \\
\label{eq:ct}
s^\mu_{CT} &=& -i C_{CT} \frac{1}{\sqrt{2} f_\pi} \cos \Theta_C \gamma^\mu \left[g_A F_{CT}^V(q^2)\gamma^5-F_\rho((q-k)^2) \right]F_{\pi}(k-q)\\
\label{eq:pif}
s^\mu_{PIF} &=& -i C_{PIF} \frac{g_A}{\sqrt{2} f_\pi} \cos \Theta_C  F_{PIF}^V(q^2)\frac{2M(2k^\mu - q )\gamma^5}{(k-q)^2-m_\pi^2}F_{\pi}(k-q)\\
\label{eq:pp}
s^\mu_{PP} &=& -i C_{PP} \frac{1}{\sqrt{2} f_\pi} \cos \Theta_C  F_{\rho}((q-k)^2)\frac{q^\mu \sh{q}}{q^2-m_\pi^2}
\end{eqnarray}

We use the convention of \cite{Hernandez:2007qq}. In our notation $f^\ast=2.16$ is the $\pi N \Delta$ coupling constant. 
This value is slightly larger than  $2.14$ used in \cite{Hernandez:2007qq}. 
With our choice free $\Delta(1232)$ width is $0.118\ \mathrm[GeV]$. 
The values of axial couplings are standard: $g_A=1.267$ and $f_\pi=93\ \mathrm{[MeV]}$. 
We use averaged masses for nucleons and pions:  $M=\frac{1}{2}(M_n+M_p)$, 
$m_\pi=\frac{1}{3}(m_{\pi^+}+m_{\pi^-}+m_{\pi^0})$ with the values given by Particle Data Group \cite{Beringer:1900zz}. 
For the $\Delta$-resonance contributions we assume $M_\Delta=1.232\ \mathrm{[GeV]}$. In the 
Delta pole ($\Delta P$) and crossed Delta pole ($C\Delta P$) amplitudes 
$P_{\alpha\beta}(p_\Delta)$ and $\Gamma_\Delta(s)$ denote the Rarita-Schwinger projection operator 
on spin-$\frac{3}{2}$ states and free $\Delta\to \pi N$ decay width. By $\Gamma^{\beta\mu}(p,q)$ we 
denote the $\Delta$ electroweak excitation vertex. 
We will give more details about the $\Delta$ propagator and decay width in the next subsection. 
The electroweak excitation vertex as
well as the set of vector and axial form factors is described in Appendix \ref{sec:delff}.
For the nucleon weak currents present in (\ref{eq:np}) and (\ref{eq:cnp}) we use the standard vector-axial 
prescription:
\begin{eqnarray}
\label{eq:nuccurr}
j^\mu_{CCN}(q)&=&V^\mu(q)-A^\mu(q)\el
V^\mu(q) &=& F_1^V(Q^2)\gmu+\frac{i}{2M}\sigma^{\mu\alpha}q_\alpha F^2_V(Q^2)\el
A^\mu(q) &=& G^A(Q^2)\left(\gal\gfiv+\frac{\sh{q}}{m_\pi^2-q^2}q^\alpha\gfiv\right).
\end{eqnarray}

From the conserved vector current (CVC) hypothesis one can also get constraints on form factors 
of contact term (CT) 
and pion-in-flight (PIF) diagrams:
\begin{eqnarray}
\label{eq:bckff}
F_{PIF}(Q^2)=F_{CT}(Q^2)=F_1^V(Q^2)
\end{eqnarray}
We choose the same nucleon form-factors as in \cite{Hernandez:2007qq}. 
Details are described in the Appendix \ref{sec:nucff}.
Our current matrix elements contain a virtual pion form factor $F_{\pi}(k-q)$ coming from the 
PIF term, where the $W$ boson interacts with a virtual pion with momentum $a=k-q$. 
The CVC forces one to include it in several other background terms. $F_\pi$ is assumed to have a monopole form:
\begin{eqnarray}
\label{eq:fpion}
F_\pi(a)=\frac{\Lambda^2_\pi-m^2_\pi}{\Lambda^2_\pi-a^2};\ \Lambda_\pi=1.25\mathrm{[GeV]}.
\end{eqnarray}
The $\rho$-meson form factor $F_\rho(a)=\frac{1}{1-a^2/m_\rho^2};\ m_{\rho}=0.7758\ \mathrm{[GeV]}.$ 
has been introduced in the PP term by the authors of \cite{Hernandez:2007qq} in order to account for the $\rho$-meson 
dominance of $\pi\pi NN$ coupling. Because of the partially conserved axial current (PCAC) hypothesis it has been also
introduced in the axial part of CT.
For each physical pion production channel there is a set of isospin Clebsch-Gordan coefficients $C_i$.
\begin{table}[htb] 
\caption{Charged current isospin coefficients of (\ref{eq:dp}-\ref{eq:pp})}
\label{tab:CG}\centering
\begin{tabular}{|c|ccccc|}
\hline
 Process & $\Delta$P & C$\Delta$P & NP & CNP & CT, PIF, PP \\
\hline
 $\nu_l+p\to l^-+\pi^++p$ & $\sqrt{3}$ & $\sqrt{1/3}$ & $0$ & $1$ & $1$ \\
 $\nu_l+n\to l^-+\pi^0+p$ & $-\sqrt{2/3}$ & $\sqrt{2/3}$ & $\sqrt{1/2}$ & $-\sqrt{1/2}$ & $-\sqrt{2}$ \\
 $\nu_l+n\to l^-+\pi^++n$ & $\sqrt{1/3}$ & $\sqrt{3}$ & $1$ & $0$ & $-1$ \\
 $\overline{\nu}_l+n\to l^++\pi^-+n$ & $\sqrt{3}$ & $\sqrt{1/3}$ & $0$ & $1$ & $1$ \\
 $\overline{\nu}_l+p\to l^++\pi^0+n$ & $\sqrt{2/3}$ & $-\sqrt{2/3}$ & $-\sqrt{1/2}$ & $\sqrt{1/2}$ & $\sqrt{2}$ \\
 $\overline{\nu}_l+p\to l^++\pi^-+p$ & $\sqrt{1/3}$ & $\sqrt{3}$ & $1$ & $0$ & $-1$ \\
\hline
\end{tabular}
\end{table}
They are listed in Tab. \ref{tab:CG}.  In the carbon cross section computations we 
sum up contributions from protons and neutrons in the incoherent way.

\subsubsection{$\Delta(1232)$ Decay Width and Propagator}
\label{sec:mediumdel}

The $\pi N\Delta$ interaction  is decribed by the Lagrangian:
\begin{eqnarray}
\label{eq:lpind}
\mathcal{L}_{\pi N \Delta}=\frac{f^\ast}{m_\pi}\opsi_\mu\vec{T}^\dag(\Dmu\vec{\phi})\psi+h.c.
\end{eqnarray}
This results in the following formula for free vacuum $\Delta\to \pi N$ decay width:
\begin{eqnarray}
\label{eq:w5a}
\Gamma^{vac.}_{\Delta\rightarrow N\pi}&=&\frac{1}{12\pi}\frac{f^{\ast 2}}{m^2_\pi} \frac{k^{3}_{cm}(E_{N, cm}+M)}{W} 
\end{eqnarray}
It is worthy to notice, that the authors of \cite{Nieves:2011pp} and \cite{Gil:1997bm} use:
\begin{eqnarray}
\label{eq:w6a}
\Gamma^{vac.}_{\Delta\rightarrow N\pi}&=&\frac{1}{12\pi}\frac{f^{\ast 2}}{m^2_\pi} \frac{k^{3}_{cm}(2M)}{W} 
\end{eqnarray}
In the above formulae $cm$ denotes the $\Delta$ center of mass frame. 

The default $\Delta$ propagator is given by:
\begin{eqnarray}
\label{eq:delprop}
G^{\alpha\beta}(p_\Delta)&=&\frac{P_{3/2}^{\alpha\beta}\left(p_\Delta,M_\Delta\right)}{p_\Delta^2 - M_\Delta^2+ i M_\Delta\Gamma_\Delta(p_\Delta^2)}\\
P_{3/2}^{\alpha\beta} \left(p_\Delta,M_\Delta\right) &=&- (\sh{p}_\Delta + M_\Delta )\left(g^{\alpha\beta}-\frac{1}{3}\gal\gbe-\frac{2}{3}\frac{p_\Delta^\alpha p_\Delta^\beta}{M_\Delta^2}+\frac{1}{3}\frac{p_\Delta^\alpha \gbe - p_\Delta^\beta\gal}{M_\Delta}\right).
\end{eqnarray}
In the above equation $P_{3/2}^{\alpha\beta}$ is the projection operator on spin-$\frac{3}{2}$ states with 
$p_\Delta$ being the $\Delta$ resonance 4-momentum and $\Gamma_\Delta$ the free resonance decay width given in 
(\ref{eq:w6a}).

\subsubsection{$\Delta$ Self-Energy}
\label{sec:deltaself}
The $\Delta(1232)$ isobar exhibits a  strongly medium-dependent behavior due to the possibility to decay into a 
pion-nucleon pair. The free resonance decay width gets decreased because of Pauli blocking. 
Assuming a uniform distribution of decay
pions in the $\Delta$ rest frame, the Pauli blocking factor is calculated to be:
\begin{eqnarray}
\label{eq:pbf}
F(p^0_\Delta,|\vec{p}_\Delta|,E_F )=\frac{p_\Delta^0 E_{N,cm} + |\vec{p}_\Delta|k_{cm}-E_FW}{|\vec{p}_\Delta|k_{cm}}
\end{eqnarray}
and 
\begin{eqnarray}
\Gamma^{vac}\to\tilde{\Gamma}=F(p^0_\Delta,|\vec{p}_\Delta|,E_F )\Gamma^{vac}(s).
\end{eqnarray}

But inside nucleus other decay channels are opened: the two- and three-nucleon absorption. 
The net effect is an overall increase of the $\Delta$ width.
\begin{eqnarray}
\Gamma_\Delta^{vac.}(s)\rightarrow 2(\frac{1}{2}\tilde{\Gamma}_\Delta+ i\Sigma_\Delta^{matter})= \tilde{\Gamma}_\Delta-2(\Im\Sigma_{1p1h1\pi}+\Im\Sigma_{2p2h}+\Im\Sigma_{3p3h} -i\Re\Sigma_\Delta)
\end{eqnarray}
In \cite{Oset:1987re} Oset parameterized this width as a functions of either the incoming pion 
kinetic energy $x=\frac{T_\pi}{m_\pi}$ or the real photon energy and the local density of nuclear matter.
We use his approach in our computations. 
It is necessary to {\it translate} the Oset results obtained in the kinematical situations of
real photon or pion scattering to the situation of virtual boson interaction.
It was assumed that the Oset functions:
\begin{eqnarray}\label{eq:deltaselfenergy}
-\Im \Sigma_\Delta=C_{1p1h1\pi}(\rho/\rho_0)^\alpha+C_{2p2h}(\rho/\rho_0)^\beta +C_{3p3h}(\rho/\rho_0)^\gamma
\end{eqnarray}
(all $C_x$ and $\alpha, \beta, \gamma$ are the functions of photon energy or pion kinetic energy) 
are in a good approximation the functions of the average invariant hadronic system mass.
The relations:
\begin{eqnarray}
\langle W^2\rangle =\left\{\begin{array}{cc}M^2+2E_\gamma\langle E_N(\rho_N ) \rangle& \quad \gamma \\ 
M^2+2E_\pi\langle E_N(\rho_N )\rangle + m_\pi^2 & \quad \pi \end{array}\right.
\end{eqnarray}
together with 
$W^2=(p_N+q)^2$ 
allow us to {\it translate} the virtual boson into one of the available parameterizations.
For the real part of self-energy we use the same prescription as in \cite{Singhy}:
\begin{eqnarray}
\label{eq:resingh}
\Re(\Sigma)=40\frac{\rho(r)}{\rho(0)} \mathrm{[MeV]}
\end{eqnarray}
This prescription neglects different renormalizations of the longitudinal and transverse $\Delta$ response 
functions in the nuclear medium, but for our purpose it is sufficient.

The main problem in using these prescriptions in model of \cite{Hernandez:2007qq} comes from the fact 
that $\Sigma_\Delta$ is calculated using nonperturbative effects not included in tree-level diagrams
of (\ref{eq:dp}-\ref{eq:pp}). 
All of them contain simple single pion interaction vertex. 
Thus we modify only the widths in denominators of $\Delta$P diagram by substituting:
\begin{eqnarray}
\label{eq:dwidthmod}
\frac{1}{p^2_\Delta-M_\Delta^2+iM_\Delta\Gamma^{vac.}(s)} \to \frac{1}{p^2_\Delta-M_\Delta^2+iM_\Delta\left[\tilde{\Gamma}-2(\Im\Sigma_\Delta -i\Re\Sigma_\Delta)\right]}
\end{eqnarray}

The many-body correction to the SPP through $\Delta$ resonance $\Im\Sigma_{1p1h1\pi}$ and cross sections for multinucleon channels connected to $\Im\Sigma_{2p2h}$ and $\Im\Sigma_{3p3h}$ can be accounted for by changing the $|\Delta P|^2$ contribution (\ref{eq:dp}).
It can be done by substituting it by a full $\Delta$ resonance production cross section:
\begin{eqnarray}\label{eq:osetdeltanucc}
\frac{d^3\sigma}{dE'd\Omega'}\hspace{-5pt}&\approx &\hspace{-5pt}\frac{G_F^2|k'|}{16\pi^5 |k|}\hspace{-5pt} \int\hspace{-5pt} dr r^2\hspace{-3pt}\int\hspace{-5pt} d^3p\frac{n_N(p)}{E(p)(M_\Delta \hspace{-3pt} + \hspace{-2pt} W )}\frac{\frac{1}{2}\tilde{\Gamma} \hspace{-2pt} -\hspace{-2pt}  \Im\Sigma_\Delta}{(W\hspace{-2pt} -\hspace{-2pt} (M_\Delta\hspace{-3pt} +\hspace{-2pt} \Re\Sigma_\Delta))^2\hspace{-3pt} + \hspace{-2pt}(\frac{1}{2}\tilde{\Gamma}\hspace{-3pt} -\hspace{-2pt} \Im\Sigma_\Delta)^2}\el & &\tr\left[\gamma^0\Gamma^{\alpha\mu^\dag}\gamma^0 P^{3/2}_{\alpha\beta}(p_\Delta )\Gamma^{\beta\nu}(\sh{p}+M)\right] L_{\mu\nu}
\end{eqnarray}
The approximation comes from the nonrelativistic expansion in the $\Delta$ propagator
$p_\Delta^2-M^2+iM_\Delta\Gamma_\Delta\approx (M_\Delta+W)(W-M_\Delta+\frac{i}{2}\Gamma_\Delta)$.
As for the isospin dependence: for electrons proton and neutron get the same factor of 1; 
for neutrinos/antineutrinos protons/neutrons get a factor of 3
because of the Clebsch-Gordan $\sqrt{3}$ in the weak $\Delta$ excitation vertex.

\section{Numerical procedures}
\label{sec:numer}
The full integration of cross-section within LDA (as given in Eqs \ref{eq:xs1} and \ref{eq:xs2}) 
even with an assumption of spherically symmetric nuclear matter distribution and on-shell nucleons would 
require performing six nested integrals. For a small $\mathcal{O}(10)$ number of integration points in each of them 
we would need to evaluate $\mathcal{O}(10^6)$ points in the numerical integration procedure 
to obtain just one point in the triple-differential cross-section. 
Thus the authors of \cite{Nieves:2011pp} assumed the nucleon momentum to be an average one in local Fermi sea, 
$\langle|\vec{p}|\rangle=\sqrt{\frac{3}{5}}p_F^N(\vec{r})$. Furthermore 
$\vec{p}$ is assumed to be orthogonal to the $(\vec{q},\vec{k})$ plane. Within this approximation
the number of nested integrals is reduced by 2:
\begin{eqnarray}
\label{eq:approx}
-\frac{1}{\pi}\Im\left(\Pi^{\mu\nu}_{1p1h1\pi}L_{\mu\nu}\right)&\approx&\sum_{iso}\hspace{-2pt}\int\hspace{-5pt}
\frac{d^3k}{\dpt}\frac{1}{2E_\pi(k) } \tr\hspace{-2pt}\left[\hspace{-2pt} A^{\mu\nu}_{1p1h1\pi}(\langle p\rangle ,q,k)
\hspace{-2pt} \right]L_{\mu\nu}\el 
&&\el
& &\hspace{-2pt}\mintt{}\frac{\left[ n_N(p)(1\hspace{-2pt} -
\hspace{-2pt} n_{N'}(p'))+ n_{N'}(p')
(1\hspace{-2pt} -\hspace{-2pt} n_{N}(p))\right]}{4E(p) E(p')}\el 
&&\el
& &\delta(E(p')- q^0 \hspace{-2pt} +\hspace{-2pt} E_\pi\hspace{-2pt} -\hspace{-2pt} E(p))
\end{eqnarray}
The integral over $d^3p$ can now be performed analytically, giving a result 
proportional to the Lindhard function. 
There are severe shortcomings of this approximation and we loose a lot of precision. 
One example is the threshold behaviour of the pion production cross section. The hadronic tensor is described by an {\it averaged} 
invariant pion-nucleon mass. Thus the physically meaningful tensor is obtained, when
\begin{eqnarray}\label{eq:effw1}
\langle W^2\rangle = M^2+2\langle E_N\rangle q^0 +q_\mu^2 \geq (M+m_\pi)^2. 
\end{eqnarray}
The above mentioned condition is important for nucleon pole (NP) diagram, for which an unphysical $W^2$ may 
give rise to a singularity at $(\langle p\rangle +q)^2=M^2$. This requires an additional cutoff in 
the acceptable kinematics, which sometimes moves up the threshold for pion production process in an artificial way.

However, the six dimensional integration can be performed using Monte Carlo techniques. 
There exist several available algorithms for that.
We have chosen the Vegas  algorithm implemented in GNU Scientific Library (GSL) for C/C++ compilers \cite{GSL}.
It is efficient enough to perform 8-dimensional total
cross section integration in a reasonable time using only $\mathcal{O}(10^5)$ points. 
This solves the threshold problem caused by
averaged hadronic tensor with averaged $W^2$. 

\begin{figure}[htb]
\centering
\[ \begin{array}{cc} \includegraphics[width=0.5\textwidth]{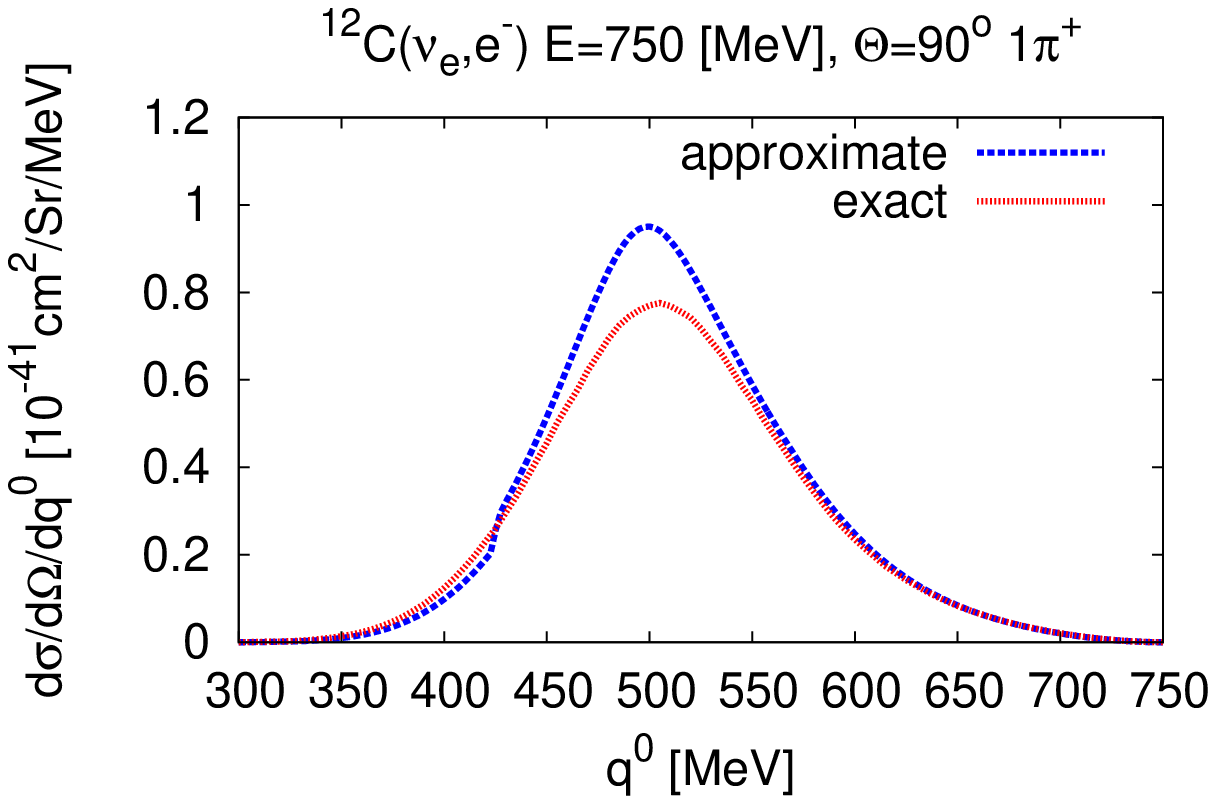} &\includegraphics[width=0.5\textwidth]{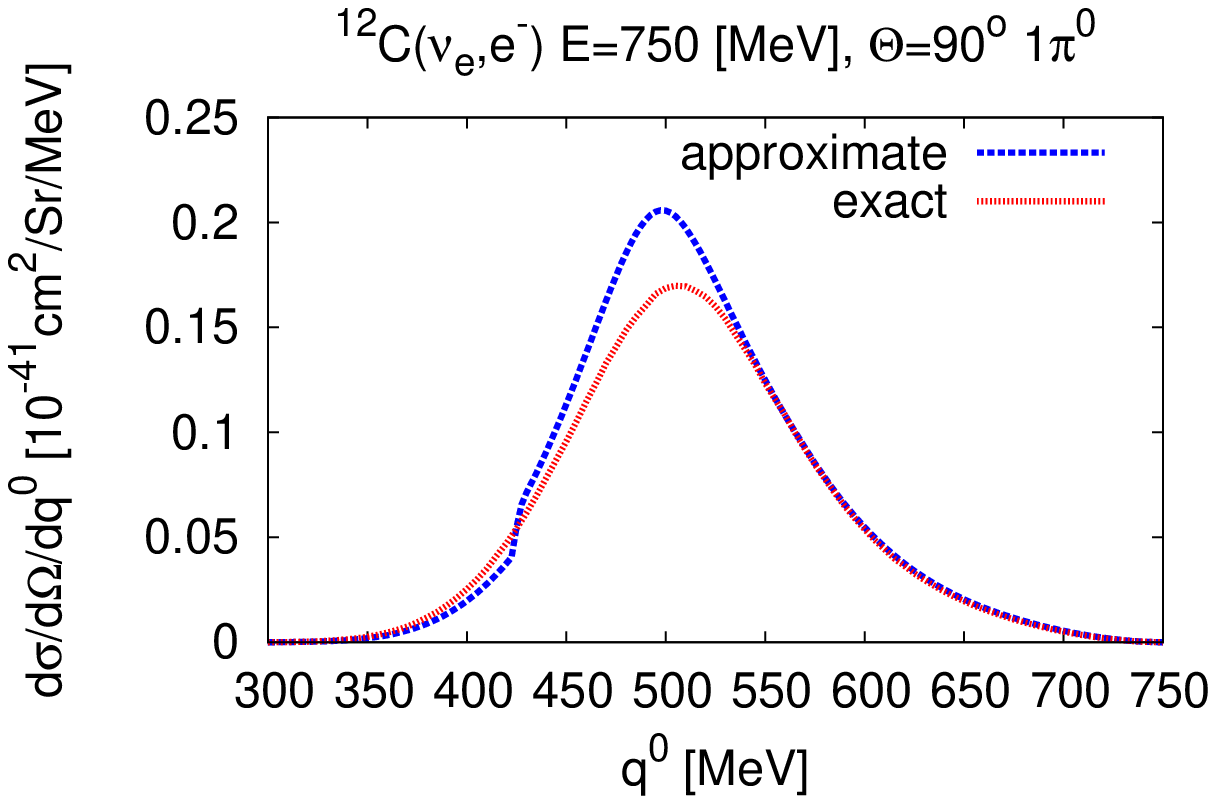}\\
\includegraphics[width=0.5\textwidth]{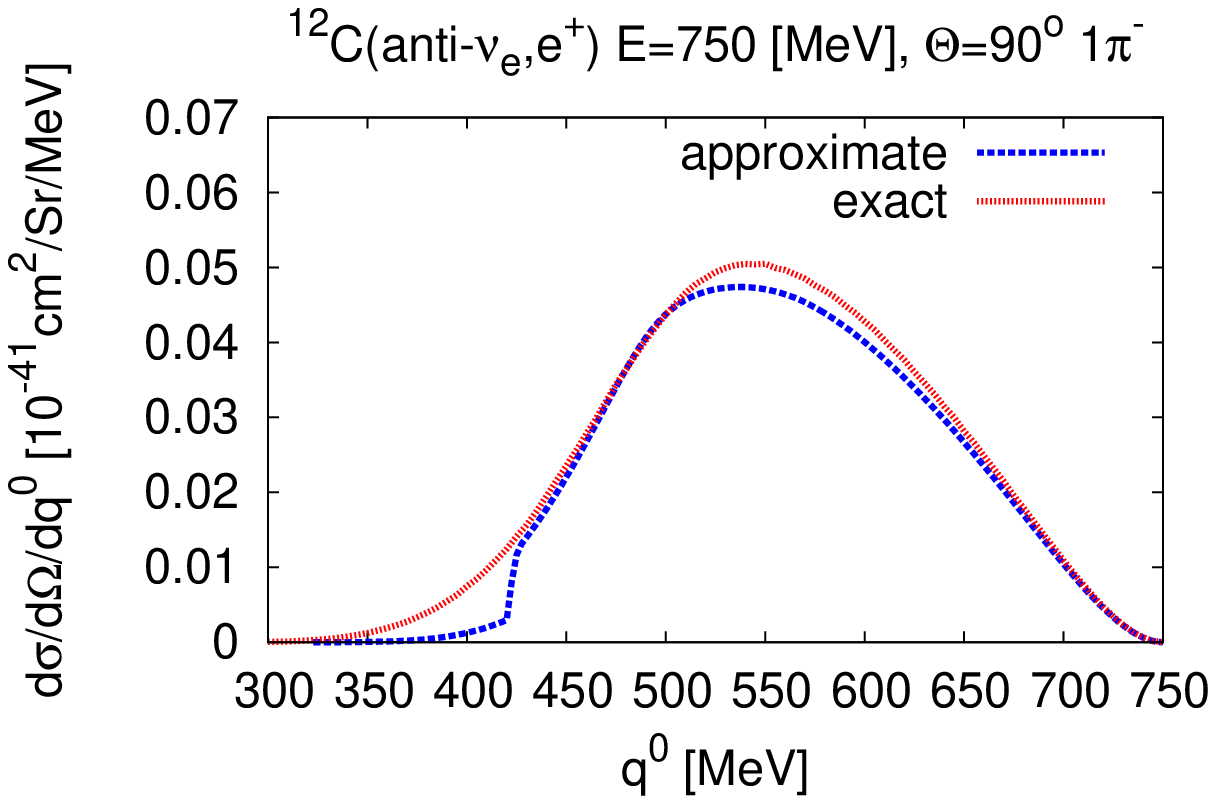} &\includegraphics[width=0.5\textwidth]{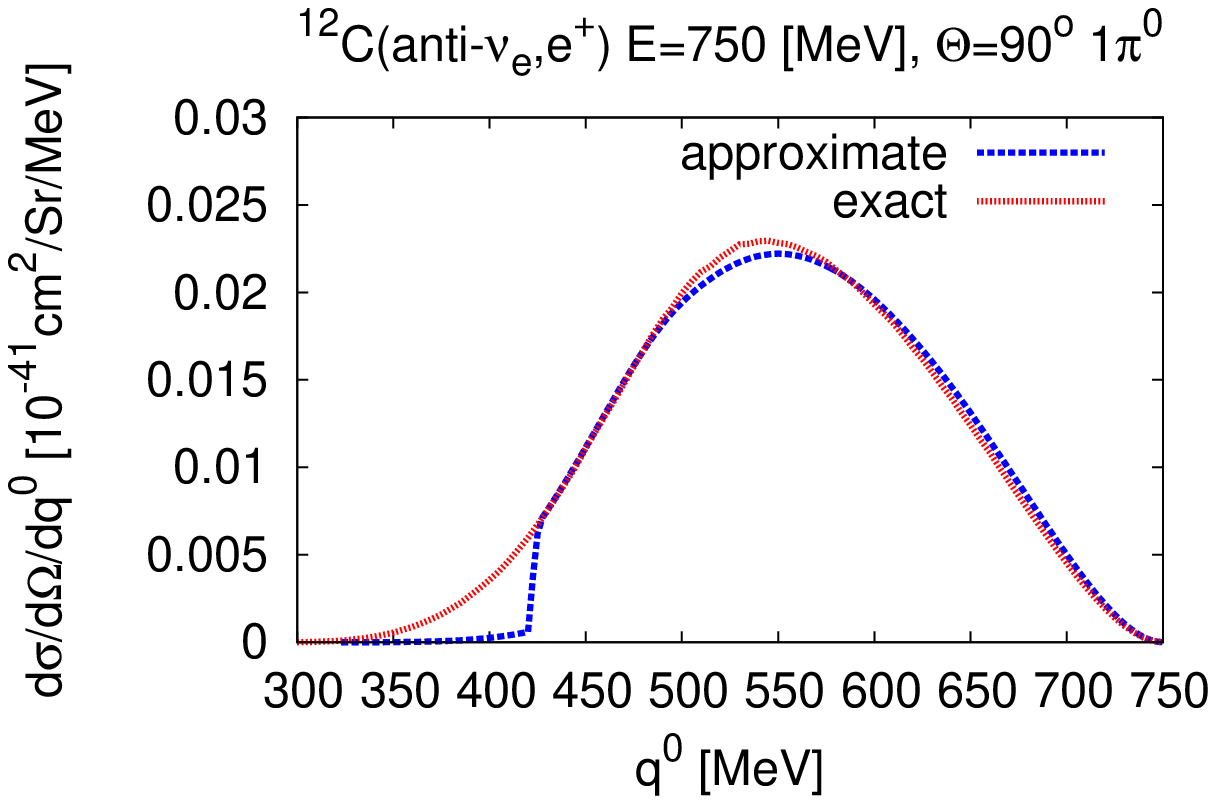} \end{array}\]
\caption{Difference between the exact cross section calculation from this paper and approximations used in 
\cite{Nieves:2011pp}.}\label{fig:diff2}
\end{figure}

In order to show the difference between the exact calculation and the approximation adopted in (\ref{eq:approx})
we calculated a sample double-differential electron neutrino cross section off carbon.
The results are shown in Fig. \ref{fig:diff2} for neutrinos (top) and for antineutrinos (bottom). 
The curves calculated using (\ref{eq:approx}) are quite 
different from those calculated without approximations. 

\begin{figure}[htb]
\centering
\[ \begin{array}{cc} \includegraphics[width=0.5\textwidth]{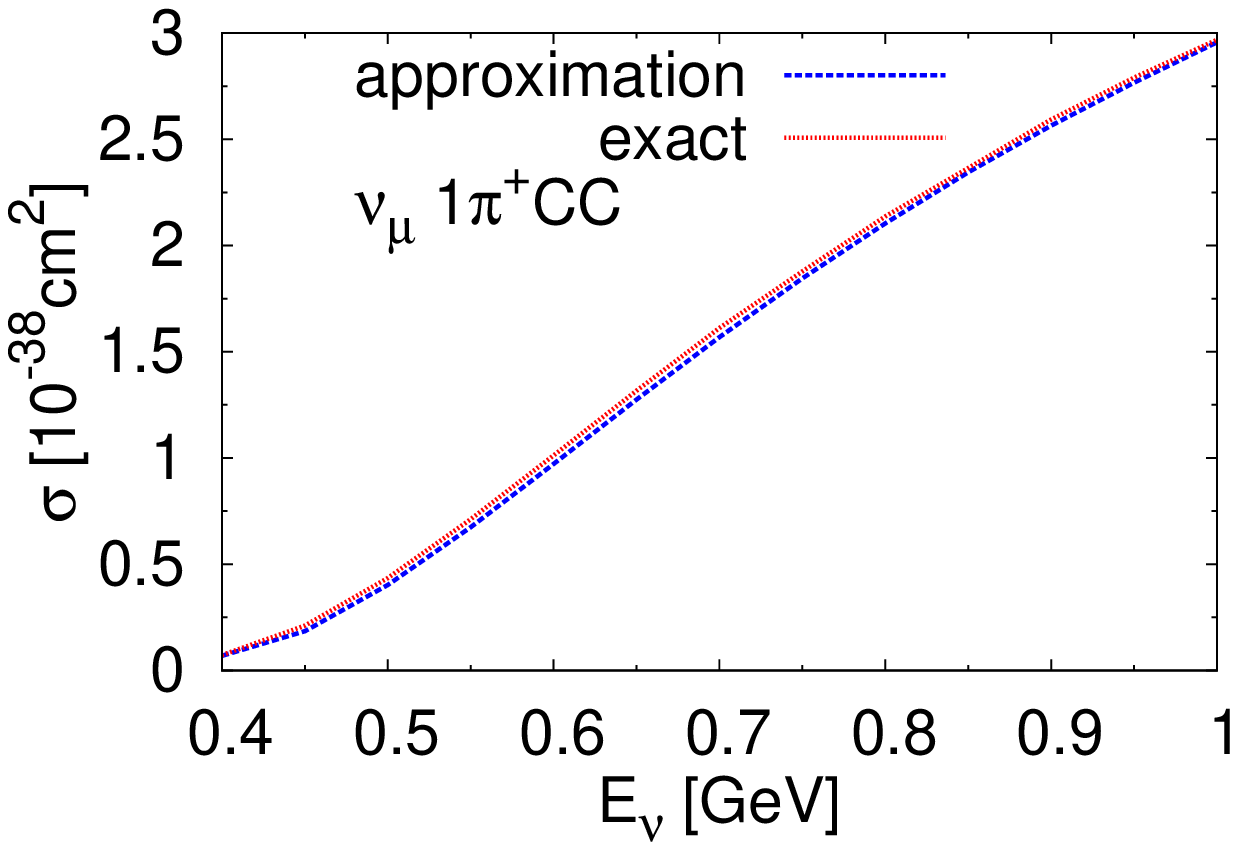} &\includegraphics[width=0.5\textwidth]{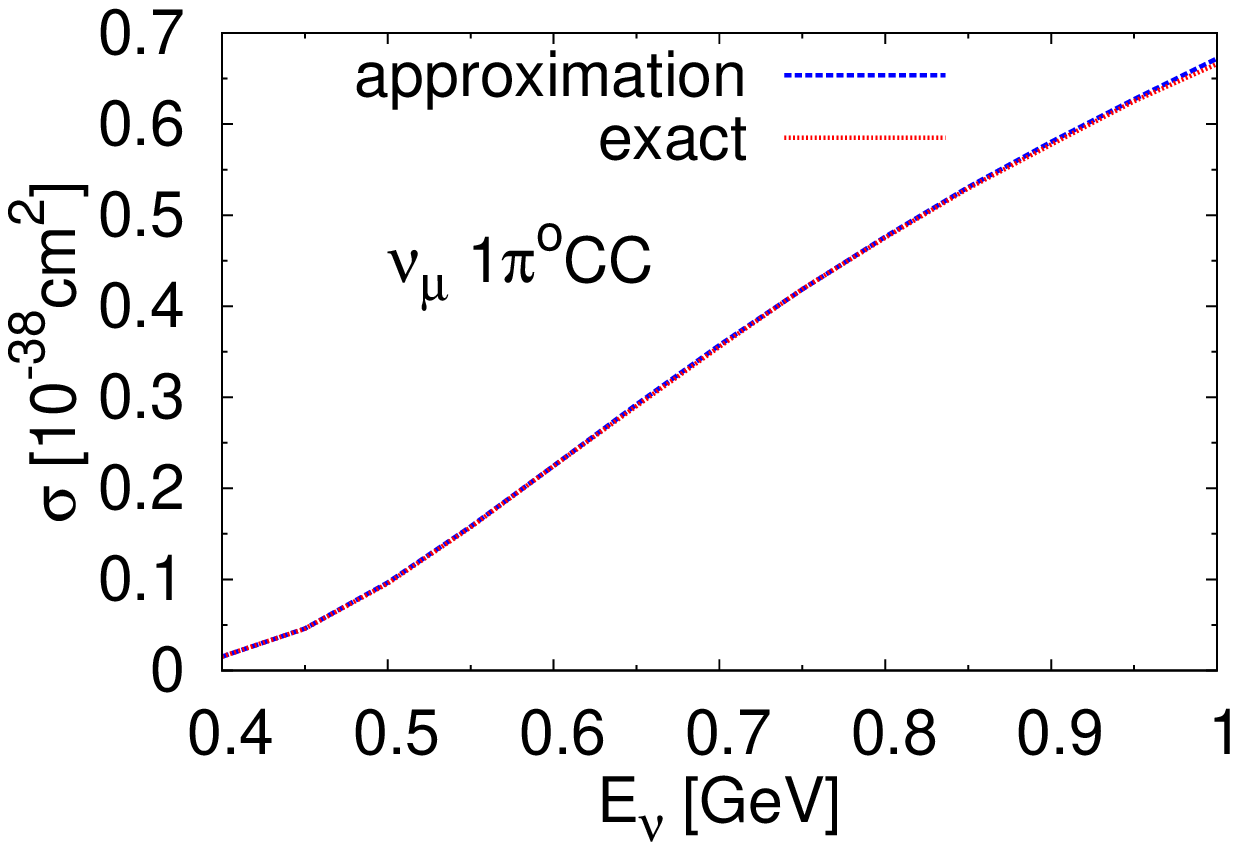} \\
\includegraphics[width=0.5\textwidth]{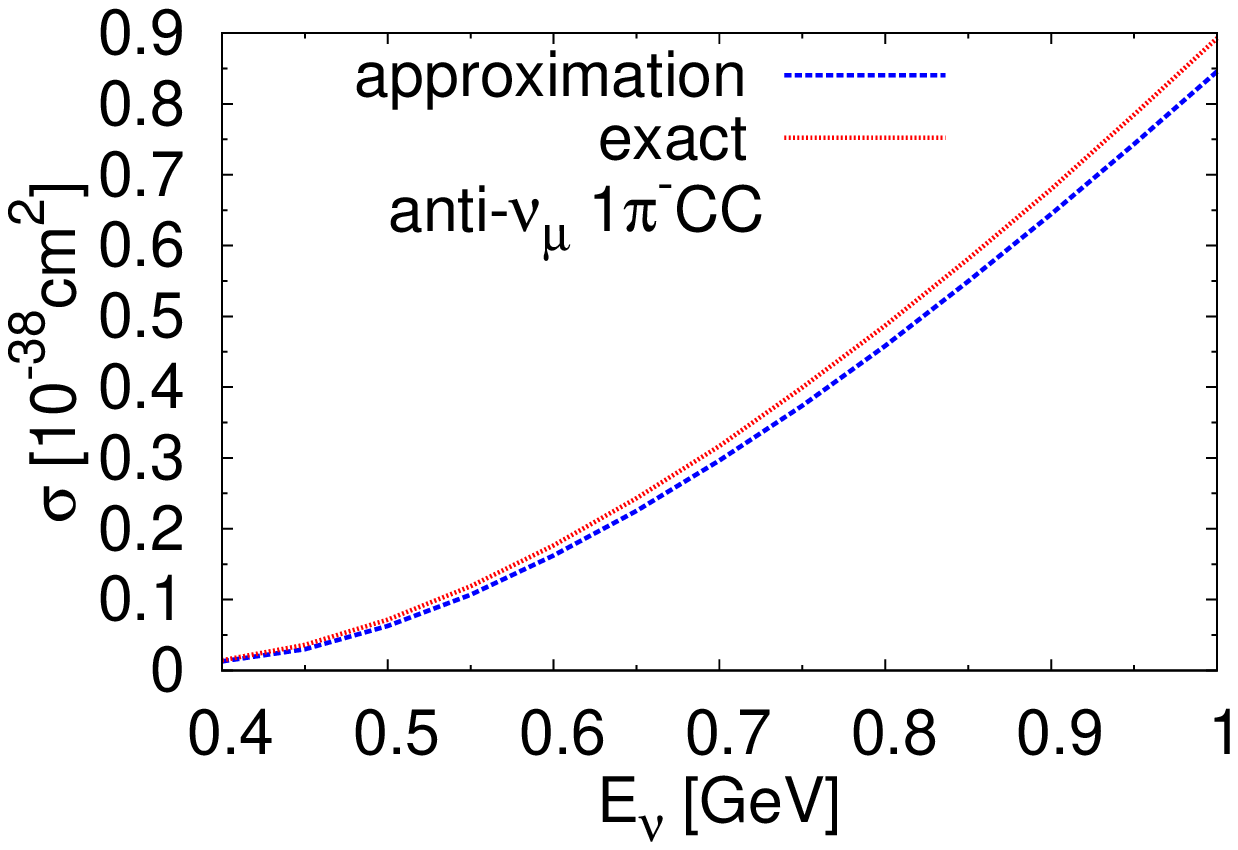} 
&\includegraphics[width=0.5\textwidth]{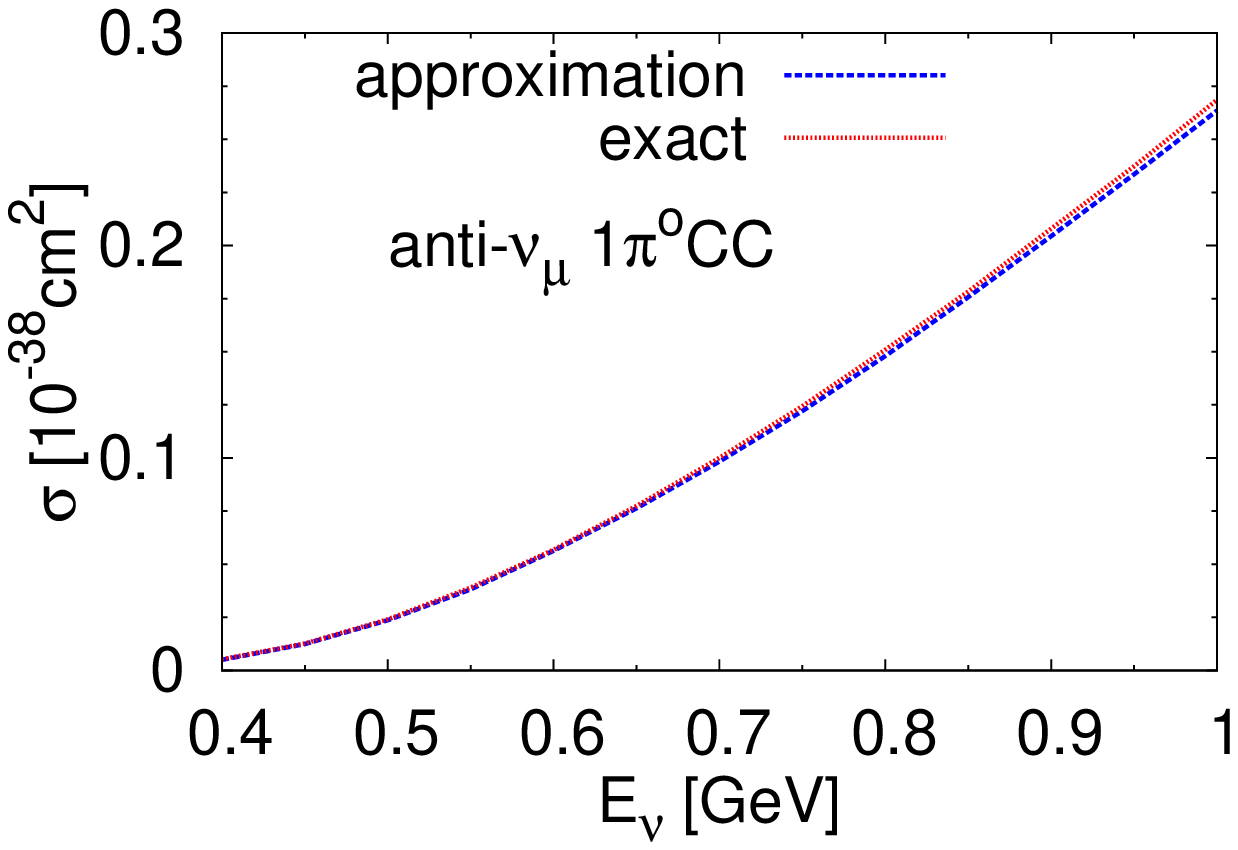}\end{array}\]
\caption{Difference between exact cross section calculation from this paper 
and approximations used in \cite{Nieves:2011pp}. 
}
\label{fig:diff3}
\end{figure}

For total cross-section both approaches: exact and approximate give similar results,
as one can see in the Fig. \ref{fig:diff3}. In the case of antineutrino charged pion production
there is a systematic difference between our calculation and approximated results, but it is rather small.
Thus we find the approximation (\ref{eq:approx}) sufficient on the level of total cross-sections.
However, in what follows we will always use the exact calculations.
% in order to reduce the numerical uncertainties of the model.

%\subsection{Differences with respect to ??}
%
%So far our model differs from the original one in several small details. We will briefly point them out once again:
%\begin{itemize}
%\item We use the Q-value correction instead of Fermi kinetic energy subtraction to account for threshold/binding effects.
%\item We use Manley-Saleski decay width (\ref{eq:deltamsw}) in the place of (\ref{eq:w6a}).
%\item We place the $\Delta P$ propagator off shell by replacing (\ref{eq:delprop}) by (\ref{eq:delprop1}).
%\item We have higher values of $f^\ast=2.16$ and $C_5^A(0)=1.2$.
%\item We use simple dipole approximation for the $Q^2$-dependence of $C_5^A$.
%\item We do not do any approximations in the numerics.
%3\end{itemize}
%

\section{Results}
\label{sec:res}
\subsection{Importance of background terms}

\begin{figure}[htb]
\centering
\includegraphics[width=0.6\textwidth]{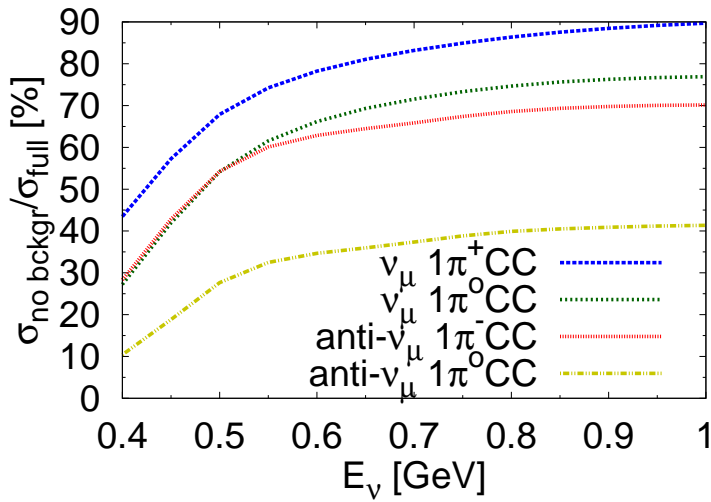} 
\caption{Ratios of the total cross sections for $\nu_{\mu}$ and $\overline{\nu}_{\mu}$ SPP reactions on carbon 
calculated with a model without the background terms to the full model of this paper.
}\label{fig:backgroundfree}
\end{figure}

Fig. \ref{fig:backgroundfree} shows importance of background terms for pion production on a set of 
6 free protons and 6 free 
neutrons. The curves describes ratios of cross sections coming 
from only Delta pole diagram to the cross section calculated 
with all the background diagrams (and their interference terms) included in computations.

We see that especially for the lower neutrino energies, below $500$~MeV, the background contribution is very important.
The background terms are more relevant for antineutrinos than for neutrinos 
and for the $\pi^0$ production than for a charged pion production.

\subsection{Importance of in-medium effects}

\begin{figure}[htb]
\centering \includegraphics[width=0.9\textwidth]{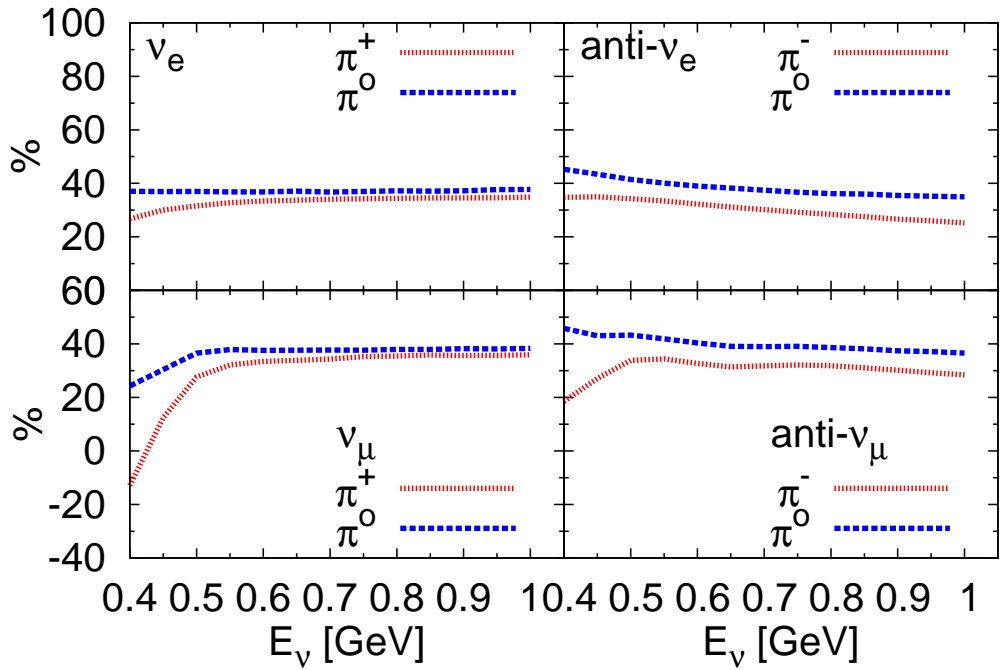}
\caption{Impact of nuclear effects on SPP off $^{12}C$. 
Plots show $(\sigma_{free}-\sigma_{medium})/\sigma_{free}\times 100\%$
}\label{fig:reduction}
\end{figure}

Fig. \ref{fig:reduction} shows an impact of the in-medium effects on the pion production. 
We plotted a relative modification
of the free nucleon cross section (six free protons and neutrons but with the background contribution included) 
caused by the in-medium effects.
In almost all of the cases the in-medium effects leads to a significant decrease of the total cross-section. 
For the electron (anti)-neutrinos $400$~MeV is far from the SPP reaction threshold and in those cases we see an 
almost constant reduction of the cross section on the level of $30-40\%$. There is 
an interesting difference in shapes between electron neutrinos 
and antineutrinos, see Fig. \ref{fig:reduction}. 
The latter exhibits a smooth drop of in-medium reduction with growing neutrino energy. 
In the case of muon neutrinos and antineutrinos near the pion production threshold  
($E_\nu <0.5\ \mathrm{[GeV]}$) the cross section is less affected by nuclear effects. For $\pi^+$ production
channel and $E_\nu =0.4\ \mathrm{[GeV]}$ it even seems to be slightly enhanced. This happens due to nucleon Fermi
motion which dominates other effects in that kinematical region. This is not the case for $\pi^0$ production by
antineutrinos. There exists a correlation between the nonresonant background contribution 
and the cross section reductiondue to in-medium effects. Shapes of the reduction ratios in neutrino $\pi^0$
and antineutrino $\pi^-$ channels are almost the same, 
so are background contribution shown in Fig. \ref{fig:backgroundfree}.
In general, the more cross section comes from background and interference terms, the smaller is the near threshold
effect.
For the larger muon neutrino/antineutrino energies 
$E>~0.6\ \mathrm{[GeV]}$ we see again an almost uniform reduction of the cross section of the order of $30\%$.

\subsection{Total cross sections}

\begin{figure}[htb]
\centering
\[ \begin{array}{cc} \includegraphics[width=0.5\textwidth]{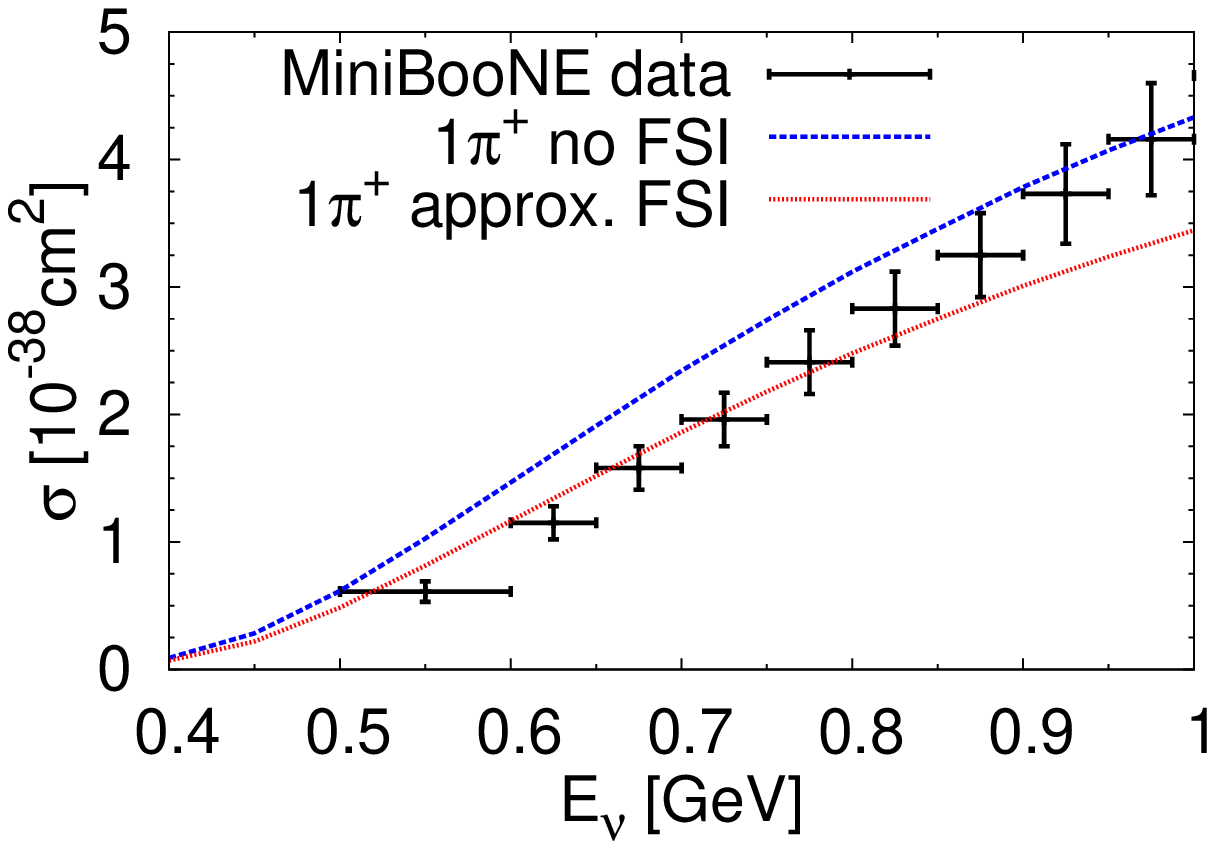} 
&\includegraphics[width=0.5\textwidth]{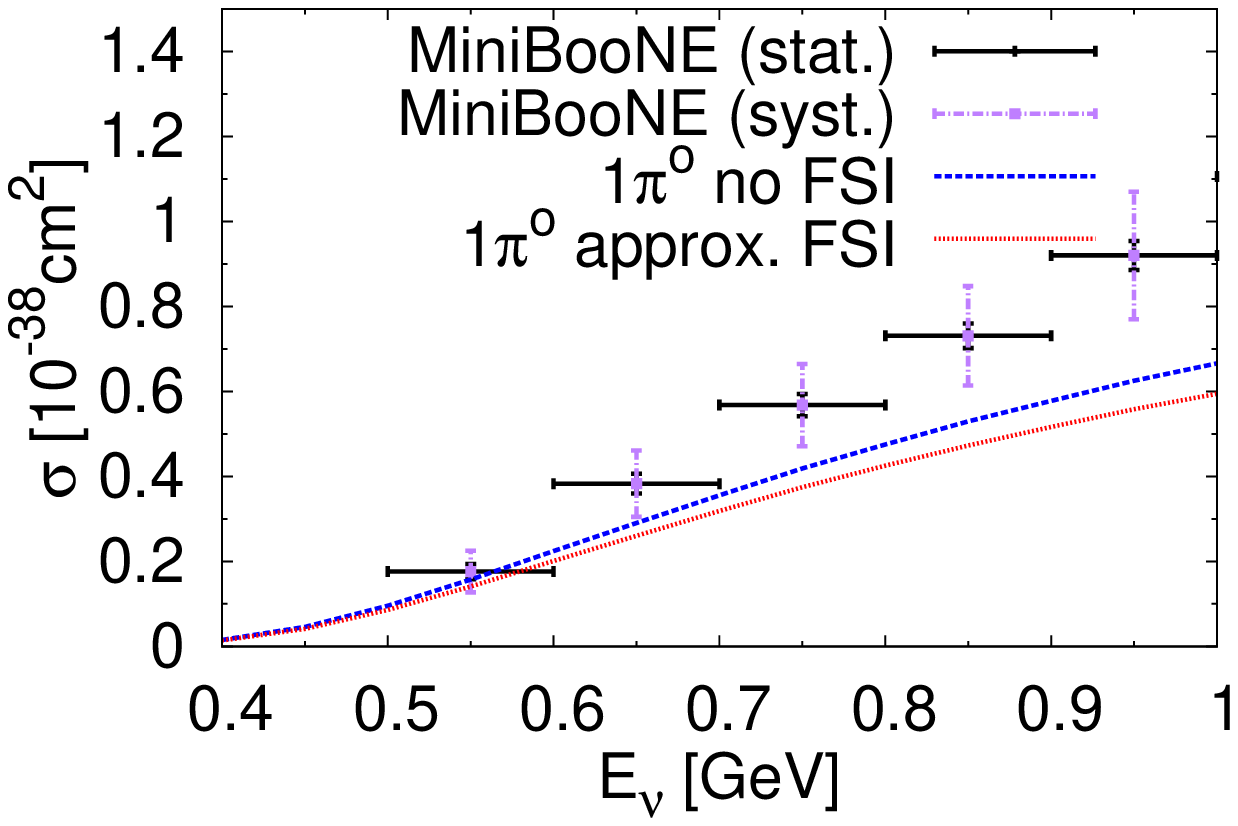} \end{array}\]
\caption{Charged and neutral pion production cross sections on $CH_2$ for full model of this paper 
plotted against the data from by 
\cite{AguilarArevalo:2010bm} and \cite{AguilarArevalo:2010xt}.}\label{fig:boonefull}
\end{figure}

We compared predictions from our model with the recent MiniBooNE pion production data.
MiniBooNE, unlike K2K, published their results in a form of absolutely normalized cross section and not as 
a ratio to CC inclusive cross sections. We performed calculations with our model of the total cross-sections on $CH_2$.
A direct comparison with the data is not straightforward because MiniBooNE reported the 
cross sections for pions in the final state after
leaving nucleus (in a case of neutrino-carbon scattering) with all the FSI effects included. 
The pion FSI effects can be
evaluated within a cascade models like those implemented in Monte Carlo event generators. Our model is not yet an 
ingredient of any MC generator and we tried to estimate an impact of FSI effects using the results of MC comparison
study published in \cite{Antonello:2009ca}. We approximate the relevant probabilities as:
\begin{eqnarray}
\label{eq:approxfsi}
P(\pi^0\to\pi^0)=67\%,\quad P(\pi^0\to\pi^+)=5\%\\
P(\pi^+\to\pi^+)=69\%,\quad P(\pi^+\to\pi^0)=5\% 
\end{eqnarray}
The results for the cross section with and without FSI are plotted in Fig. \ref{fig:boonefull}.

In the case of charged pion production we obtained a quite good agreement with the data up to 
the neutrino energy of around 
$0.8$~[GeV].
In the case of charged-current $1\pi^0$ production both free and in-medium
cross sections with our model are too small, and the discrepancy becomes larger with increasing neutrino energy.
FSI introduce large modifications for the $\pi^+$ channel. In the $\pi^0$ channel an 
effect of absorption of $\pi^0$ is partially compensated
by a fraction of initial $\pi^+$ events, that end up as $\pi^0$ due to charge exchange reaction inside nucleus.
It is important to point out that in the case of CC$\pi^0$ reaction 
also the computations of other theoretical groups give results well below
the measured cross section \cite{mosel}.

\subsection{Ratios of muon to electron (anti-)neutrino cross sections}

In neutrino oscillation appearance experiments it is very important to calculate precisely 
ratios of muon and electron
neutrino cross sections. Even in a presence of a near detector and with full understanding of initial muon neutrino flux
a  good knowledge of the ratios (and their dependence on neutrino energy) is crucial for a correct
identification of the oscillation signal.

\begin{figure}[htb]
\centering \includegraphics[width=1.0\textwidth]{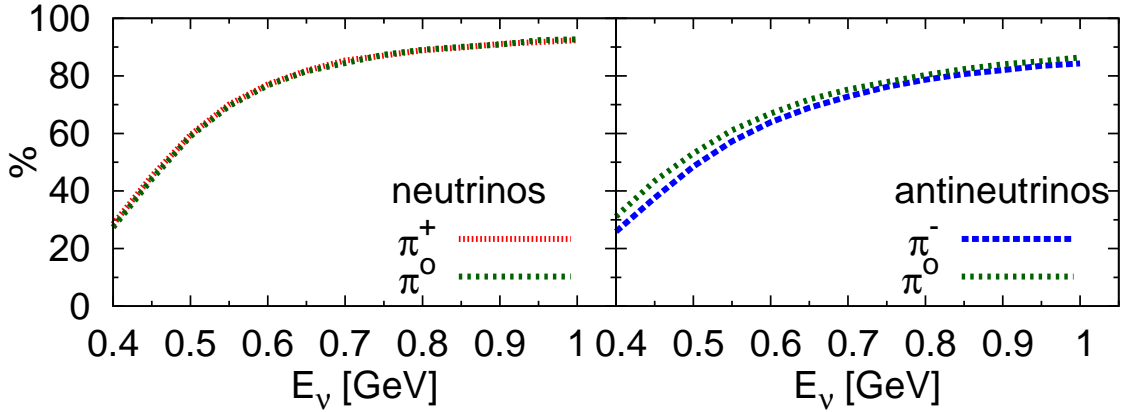}
\caption{Ratios of muon to electron (anti) neutrino total SPP cross sections on $^{12}C$ 
for the full model of this paper.}\label{fig:ratios}
\end{figure}

In Fig. \ref{fig:ratios} we see that the ratios calculated with the complete model are 
slowly increasing functions of the neutrino energy. In the case of antineutrinos
there is a small difference between $\pi^-$ and $\pi^0$ production: in the first case the ratio is slightly
lower.
On the contrary, we obtain almost the same ratios both for $\pi^+$ and $\pi^0$ production by neutrinos.

It is important to know how well the ratios are calculated when much simpler models are used, 
which is a case in MC event generators.

\begin{figure}[htb]
\centering \includegraphics[width=1.0\textwidth]{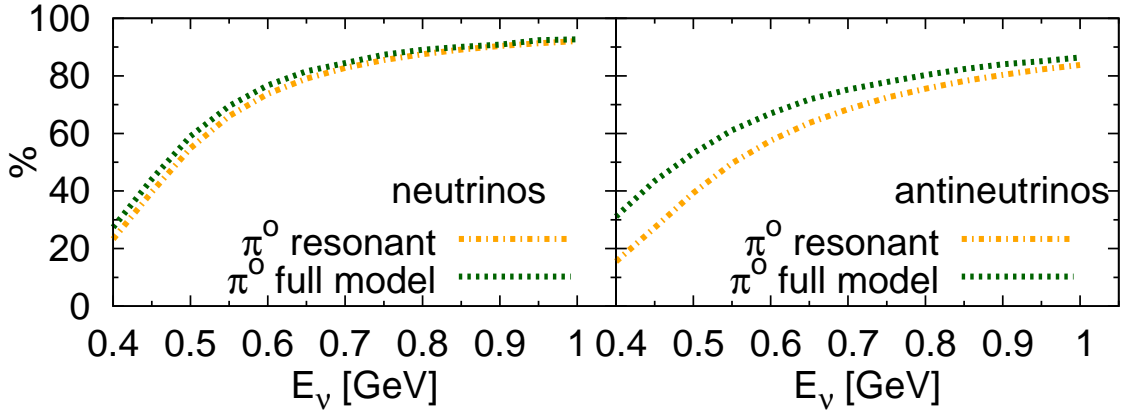}
\caption{Ratios of muon to electron (anti-)neutrino total SPP cross sections on $^{12}C$ 
for the full model of this paper 
and for the resonant SPP only (without the background terms).}\label{fig:ratios1}
\end{figure}

Fig. \ref{fig:ratios1} shows an impact of the background terms on the $\pi^0$ production ratios.
We compared two situations: the full model and the model without background contributions.
We see that the results are significantly different only in the case of antineutrinos. For lower neutrino energies
one obtains much smaller ratios while using pure resonant SPP mechanism. For neutrinos
these differences are negligible.

\begin{figure}[htb]
\centering \includegraphics[width=1.0\textwidth]{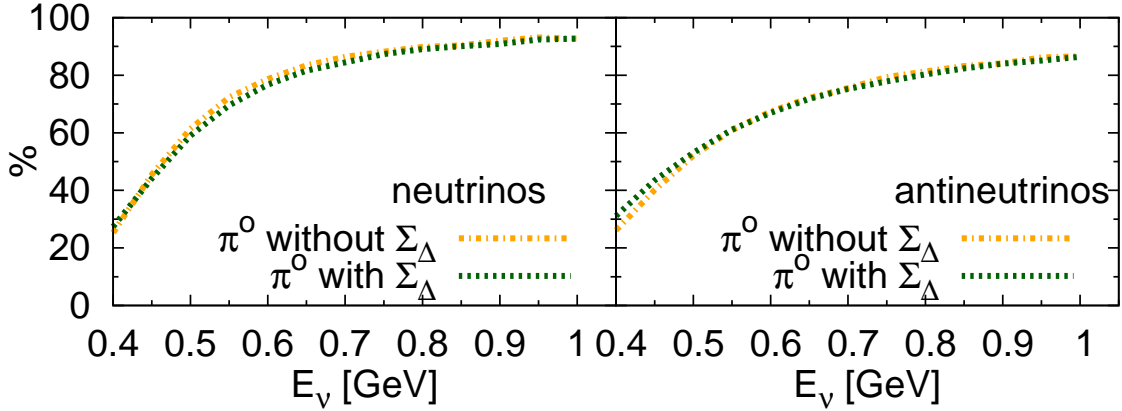}
\caption{Ratios of muon to electron (anti) neutrino total SPP cross sections on $^{12}C$ 
for the full model of this paper with and without the $\Delta$ selfenergy $\Sigma_\Delta$.}\label{fig:ratios2}
\end{figure}

 Fig \ref{fig:ratios2} show an impact of $\Delta$ self-energy on the ratios. 
We compared two situations: the full model and the model
without $\Delta$ self-energy. We see that the negligence of the $\Delta$ self-energy has
almost no impact on the considered observable.
We conclude, that in order to describe well the anti-muon to anti-electron neutrino cross section ratio
it is important to include the nonresonant background, but not necessarly the $\Delta$ self-energy.

\subsection{Pionless $\Delta$ decays}

An interesting feature of the model we discuss is that we obtain a contribution 
to the cross section coming from pionless
$\Delta$ decays. This is a part of the meson exchange current (MEC) cross section which has recently attracted
a lot of attention \cite{mec}. There is a lot of evidence that 
MEC mechanism is responsible for a large CCQE axial mass measurement
reported by the MiniBooNE collaboration. Theoretical microscopic computations always 
include pionless $\Delta$ decays as 
a part of the calculated effect. Some MC event generators (NEUT, NUANCE) 
assume a constant fraction of the pionless
$\Delta$ decays and we find it interesting to check how well this assumption is satisfied in our model.

\begin{figure}[htb]
\centering \includegraphics[width=0.6\textwidth]{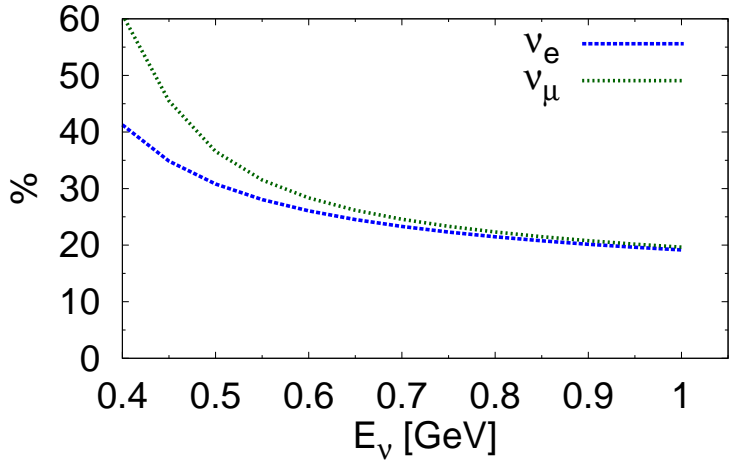}
\caption{ Fraction of the pionless $\Delta$ decays to the resonant SPP production
cross section $(\sigma_{pionless\ \Delta})/\sigma_{SPP\ res.}\times 100\%$ in $^{12}C$
for $\nu_e$ and $\nu_{\mu}$. 
}\label{fig:pionless}
\end{figure}

The fractions of the pionless decays and their dependence
on the neutrino energy and species are shown in Fig. \ref{fig:pionless}. There is no difference between neutrinos and 
antineutrinos, because we include only
the $np-nh$ mechanism coming from the resonant diagrams. The fraction of pionless $\Delta$ decays is very large
for the energies below $500$~MeV. 
For the larger energies it exhibits a smooth energy dependence, dropping down to $~20\%$ at $E_\nu=1\ \mathrm{[GeV]}$. 
It is clear that for experiments with a large 
fraction of neutrinos with energies below $1\ \mathrm{[GeV]}$ one can not consider the investigated quantity
to be constant.

\begin{figure}[htb]
\centering
\[ \begin{array}{cc} \includegraphics[width=0.5\textwidth]{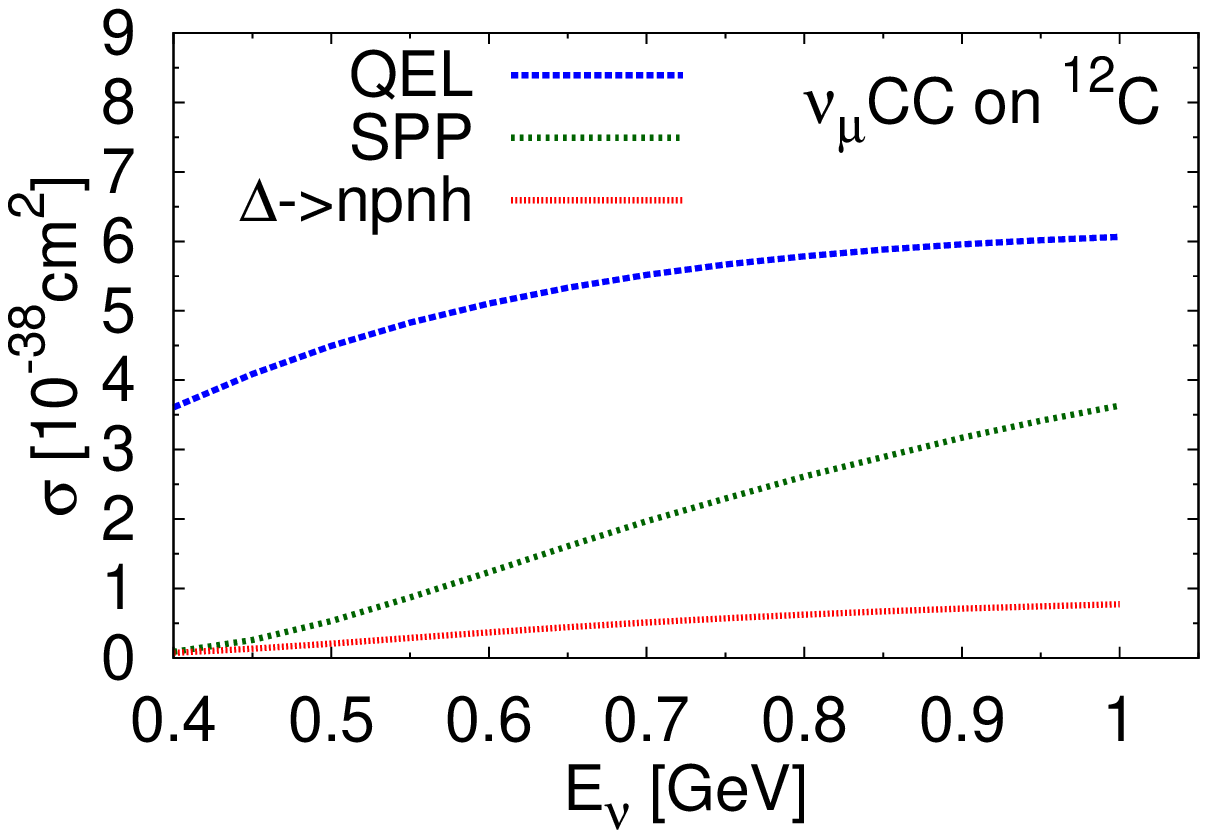} 
&\includegraphics[width=0.5\textwidth]{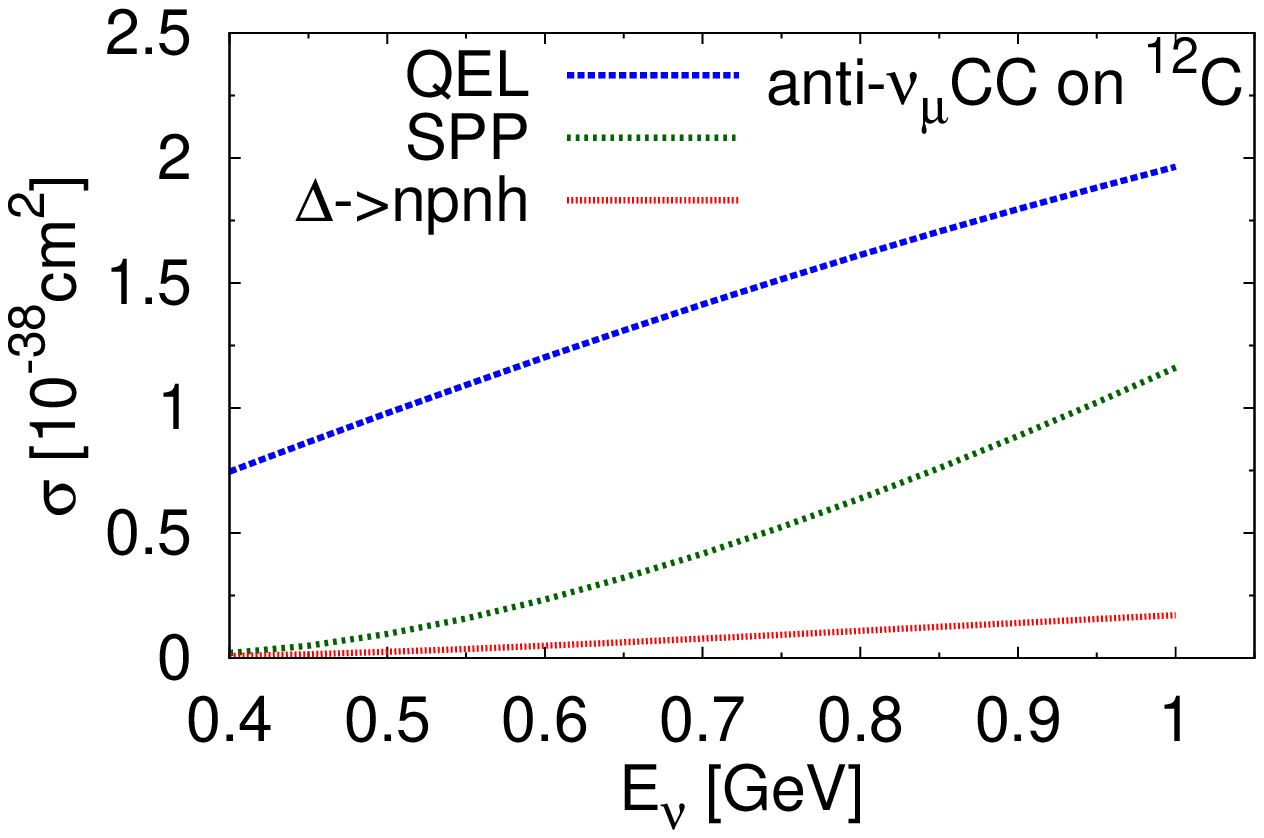} \end{array}\]
\caption{ Total charge current cross sections on $^{12}C$ for: 
quasielastic scattering, SPP, and pionless $\Delta$ decay.}
\label{fig:fullcross}
\end{figure}

The total pionless $\Delta$ decay cross section may be treated as a lower bound for the $np-nh$ contribution. One has to 
keep in mind, that there are many more sources of
$np-nh$ final states, which can be built from diagrams (\ref{eq:dp}-\ref{eq:pp}) but are not considered
in this paper. The total cross section coming from $\Delta\to np-nh$ decays can be seen in  Fig. \ref{fig:fullcross}.
The charge current quasielastic (CCQE) contribution has been calculated with NuWro neutrino event generator
\cite{nuwro} with $M_A=1.05$~[GeV]. The $np-nh$ contribution coming
from pionless $\Delta$ decays is small compared to CCQE and SPP dynamics (around 10-15\% of the first),
but it is non-negligible. Alone, it cannot explain MiniBooNE's large axial mass measurement because 
that this is only a part of the $np-nh$ cross section.

\section{Discussion}\label{sec:discuss}
It is not east to understand why the $\Delta$ self-energy leads to the 
substantial reduction of the cross sections.
As explained in the introduction the in-medium $\Delta$ spectral function was included only in the $\Delta P$ 
diagram. It is difficult to conclude, whether the cross section reduction is a genuine 
physical effect or rather an artifact
of the approximate treatment of background terms and in-medium $\Delta$ self-energy.
The pure background contribution (36 out of 49 combinations from (\ref{eq:dp})-(\ref{eq:pp})) 
is not affected by the presence of nuclear matter.
In the $\Delta$-background interference terms (12 combinations) the in-medium effects enter only through 
$\Delta P$ diagram, thus are included only partially. A complete in-medium dressing
is present only in the pure $\Delta$ contribution, reducing its size significantly. 
A verification of the model prediction can come only from the evaluation of the
non-perturbative in-medium effects for all the genuine amplitudes (28 independent terms) 
which is a very difficult task to achieve. 

We arrived at a reasonable agreement with the MiniBooNE CC$\pi^+$ production but our model underestimates
CC$\pi^0$ cross section. There can be several reasons for that. The first one can be approximations 
discussed in the previous paragraph. For neutrino energies around $1$~GeV
one should include also contributions from 
heavier resonances absent in our computations. 
It is also possible that the $1\pi2p2h$ process contributes with a larger cross section than it is generally
expected. An apparent excess of the CC$1\pi^0$ cross section with respect to several theoretical models
predictions is an interesting topic for the further research. 

We investigated a possible impact on predictions from the model coming from different descriptions of the $\Delta$  
resonance width. For example the authors of \cite{Manley:1992yb} use:
\begin{eqnarray}
\label{eq:deltamsw}
\Gamma_{M-S}(W)&=&118[MeV]\cdot\frac{\rho_{\Delta\rightarrow \pi N}(W)}{\rho_{\Delta\rightarrow \pi N}(M_\Delta )}\el
		\rho_{\Delta\rightarrow \pi N}(W)&=&\frac{k_{cm}}{W}\frac{k^2_{cm}R^2}{1+k^2_{cm}R^2}\,\ R=1[fm]
\end{eqnarray}
The term $\frac{k^2_{cm}R^2}{1+k^2_{cm}R^2}$ is a so-called Blatt-Weisskopf centrifugal barrier. 
In this manner one accounts for the phenomenological knowledge about decay $\pi N$ system angular momentum, 
which is absent in the Lagrangian (\ref{eq:lpind}). Furthermore, $\Delta(1232)$ is not a stable particle. 
One can account partially for the off-shell effects by replacing the propagator in DP term (\ref{eq:delprop}) by
\begin{eqnarray}
\label{eq:delprop1}
\tilde{G}^{\alpha\beta}(p_\Delta)&=&\frac{\tilde{P}_{3/2}^{\alpha\beta}(p_\Delta)}{p_\Delta^2 - M_\Delta^2+ i W\Gamma_\Delta(p_\Delta^2)}=-\frac{(\sh{p}_\Delta + W)}{p_\Delta^2 - M_\Delta^2+ i W\Gamma_\Delta(p_\Delta^2)}\times\el&\times& \left(g^{\alpha\beta}-\frac{1}{3}\gal\gbe-\frac{2}{3}\frac{p_\Delta^\alpha p_\Delta^\beta}{W^2}+\frac{1}{3}\frac{p_\Delta^\alpha \gbe - p_\Delta^\beta\gal}{W}\right).
\end{eqnarray}
This convention is used by \cite{Leitner:2009phd} together with the Manley-Saleski decay width. 
Thus while we use (\ref{eq:deltamsw}) we also replace (\ref{eq:delprop}) by (\ref{eq:delprop1}). In order to stay consistent, after changing the width (\ref{eq:w5a}) with (\ref{eq:deltamsw}) in 
(\ref{eq:delprop1}) one has to multiply the whole expression by $\sqrt{\frac{\Gamma_{M-S}(W)}{\Gamma_\Delta(W)}}$. 
It will compensate for the fact, that our current has a decay vertex defined by (\ref{eq:lpind}) in the numerator, 
which leads to the width (\ref{eq:w5a}).

\begin{figure}[htb]
\centering \includegraphics[width=1.0\textwidth]{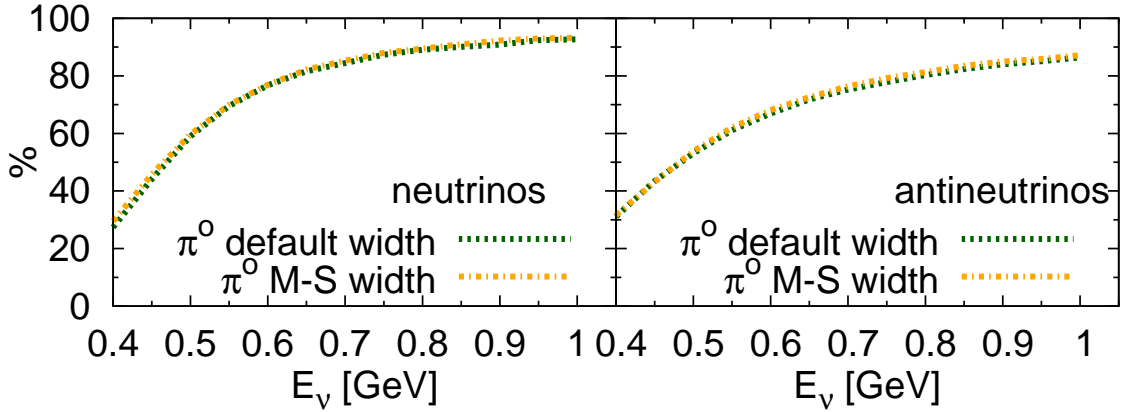}
\caption{Ratios of muon to electron (anti) neutrino total SPP cross sections on $^{12}C$ for the full model of 
this paper 
with $\Delta$ width described by (\ref{eq:w5a}) and (\ref{eq:deltamsw}).}\label{fig:ratios3}
\end{figure}

On the level of total cross sections we find the difference between two $\Delta\pi N$ decay descriptions 
negligible. This is illustrated in the Fig. \ref{fig:ratios3} where 
we plot again the muon to electron (anti-)neutrino total $\pi^0$ 
production cross section ratios and we compare the default and the 
Manley-Saleski $\Delta$ description. Both descriptions lead to almost
identical results.

Finally, we would like to address the question: how much does the numerical approximation (\ref{eq:approx}) 
affect muon to electron neutrino
cross section rates. We have already shown, that the exact integration does not change 
much total cross sections.
\begin{figure}[htb]
\centering \includegraphics[width=0.6\textwidth]{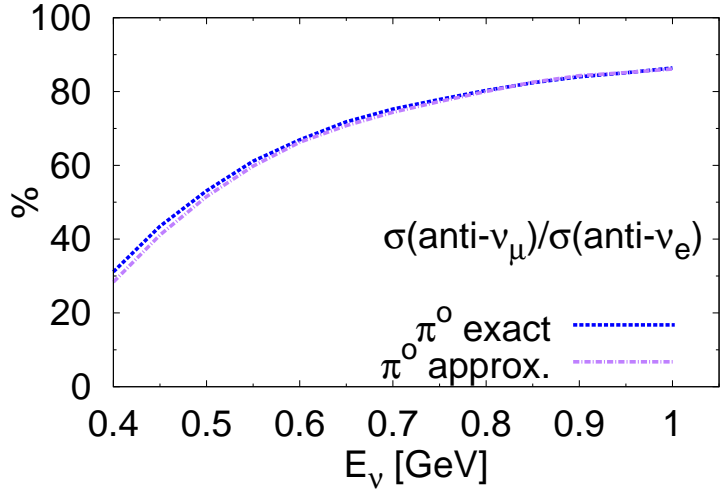}
\caption{Ratios of $\overline{\nu}_{\mu}$ to $\overline{\nu}_e$ neutrino CC$\pi^0$ SPP 
cross sections on $^{12}C$ calculated with the full model of this paper and with 
approximations used in \cite{Hernandez:2007qq}.}\label{fig:ratios4}
\end{figure}
This is illustrated in Fig. \ref{fig:ratios4}, where we have plotted 
$\overline{\nu}_\mu/\overline{\nu}_e$ $1\pi^0CC$ cross section ratios.
Differences are only for energies $E_\nu<550$~[MeV] and at $E\nu=500$~[MeV] it is about 
4.3\%.

\section{Acknowledgements}
J\.Z would like to thank L. Alvarez-Ruso, K. Graczyk and J. Nieves,  for many fruitful discussions.
This work was sponsored by grants: 4525/PB/IFT/11 (DEC-2011/01/N/ST2/03224), 4433/PB/IFT/10 (N N202 368439),   
4574/PB/IFT/12 
(UMO-2011/01/M/ST2/02578) 

\appendix

\section{Notation and conventions}
We adapt the conventions from Bjorken-Drell ($g^{\mu\nu}=(+,-,-,-)$ etc.), the only difference is in the 
Dirac spinor normalization:
\begin{eqnarray}
\sum_s \ups{p}{s}\oups{p}{s}=(\sh{p}+m)
\end{eqnarray}
which is convenient for our calculations.

\section{Nuclear Matter Density Parameterization}
\label{sec:ldapar}
We took the harmonic oscillator density profiles from \cite{DeJager:1987qc}:
\begin{eqnarray}
\label{eq:hodens}
\rho(\vec{r})=\rho_0\left(1+a\left(r/R\right)^2\right)\exp\left[-\left(r/R\right)^2\right]
\end{eqnarray}
with corrections to parameters $a$ and $R$ calculated in \cite{GarciaRecio:1991wk}. These parameters
are slightly different for protons and neutrons. The local Fermi momentum is calculated from relation 
$k_F^N(\vec r)=(3\pi^2\rho(\vec{r})^N)^{\frac{1}{3}}$.
Authors of \cite{Nieves:2011pp} subtract Fermi kinetic energy from nucleons inside medium $E(p)\to E(p)-T_F$.
In this manner they account for the binding effects.

\section{Nucleon Form Factors}
\label{sec:nucff}
The isospin symmetry relates the vector form factors to the electromagnetic ones:
\begin{eqnarray}
\label{eq:vecffn1}
F_i^V(Q^2)=F_i^p(Q^2) - F_i^n(Q^2).
\end{eqnarray}
For the electromagnetic form factors we use the parameterization of Galster \textit{et al.} \cite{Galster:1971kv}:
\begin{eqnarray}
\label{eq:galster}
F_1^N(Q^2)=\frac{G_E^N+\tau G_M^N(Q^2)}{1+\tau};\ F_2^N(Q^2)=\frac{G_M^N(Q^2)-G_E^N(Q^2)}{1+\tau}\el
G_E^p(Q^2)=\frac{G^p_M(Q^2)}{\mu_p}=\frac{G^n_M(Q^2)}{\mu_n}=-(1+\lambda_n\tau)\frac{G_E^n(Q^2)}{\mu_n\tau}=\frac{1}{(1+\frac{Q^2}{M_D^2})^2}
\end{eqnarray}
with $\mu_p=2.792.847$, $\mu_n=1.913043$, $\lambda_n=5.6$, $\tau=\frac{Q^2}{4M^2}$ and $M_D=0.843\ \mathrm{[GeV]}$. We assume the axial nucleon form factor in a dipole form:
\begin{eqnarray}
\label{eq:axialny}
G_A(Q^2)=\frac{g_A}{(1+\frac{Q^2}{M_A^2})^2};\ M_A=1.05 [GeV]
\end{eqnarray}
with $g_A=1.267$.

\section{$\Delta(1232)$ Form Factors}
\label{sec:delff}
The most general electroweak $\Delta$ excitation vertex is given by:
\begin{eqnarray}
\label{eq:delver}
\Gamma^{\alpha\mu}(p,q)&=& \left[V^{\alpha\mu}_{3/2}- A^{\alpha\mu}_{3/2}\right] = \el &=&\left[\frac{C^V_3}{M}(g^{\alpha\mu}\sh{q}  -  q^\alpha\gamma^\mu ) \hspace{-2pt} + \hspace{-2pt} \frac{C^V_4}{M^2}(g^{\alpha\mu}q \hspace{-2pt} \cdot \hspace{-2pt} (p \hspace{-2pt} + \hspace{-2pt} q) \hspace{-2pt} - \hspace{-2pt}q^\alpha (p \hspace{-2pt} + \hspace{-2pt} q)^\mu) \hspace{-2pt} +\right.\el&+&\left. \hspace{-2pt} \frac{C^V_5}{M^2}(g^{\alpha\mu}q \hspace{-2pt} \cdot \hspace{-2pt} p \hspace{-2pt}- \hspace{-2pt} q^\alpha p^\mu ) \hspace{-2pt} + \hspace{-2pt} g^{\alpha\mu} C_6^V \hspace{-2pt} \right] \hspace{-3pt}\gamma^5\hspace{-2pt}+\el
&+&\left[\hspace{-2pt}\frac{C^A_3}{M}(g^{\alpha\mu}\sh{q} \hspace{-2pt} - \hspace{-2pt} q^\alpha\gamma^\mu ) \hspace{-2pt} + \hspace{-2pt} \frac{C^A_4}{M^2}(g^{\alpha\mu}q \hspace{-2pt} \cdot \hspace{-2pt} (p \hspace{-2pt} + \hspace{-2pt} q) \hspace{-2pt} - \hspace{-2pt}q^\alpha (p \hspace{-2pt} + \hspace{-2pt} q)^\mu)\hspace{-2pt} +\right.\el &+&\left.  C_5^A g^{\alpha\mu} \hspace{-2pt} + \hspace{-2pt} \frac{C^A_6}{M^2}q^\alpha q^\mu\hspace{-2pt} \right].
\end{eqnarray}
The $C_i^V$ and $C_i^A$ vector and axial form factors determine $WN\Delta$ transition. 
For the vector form-factor set we use the  parameterization of \cite{Lalakulich:2006sw}:
\begin{eqnarray}
\label{eq:deltaffv}
C_3^V&=&\frac{2.13}{(1+Q^2/M_V^2)^2}\times\frac{1}{1+Q^2/4M_V^2}\el
C_4^V&=&\frac{-1.51}{(1+Q^2/M_V^2)^2}\times\frac{1}{1+Q^2/4M_V^2}\el
C_5^V&=&\frac{0.48}{(1+Q^2/M_V^2)^2}\times\frac{1}{1+Q^2/0.776M_V^2}\el
\end{eqnarray}
with $M_V=0.84\ \mathrm{[GeV]}$. The CVC implies that $C_6^V=0$. 
The axial part is dominated by $C_5^A$ contribution. We use a dipole approximation, in which
\begin{eqnarray}
\label{eq:c5a}
C_5^A(Q^2)=\frac{C_5^A(0)}{(1+Q^2/M_{A\Delta}^2)^2}.
\end{eqnarray}
We use a default value of the $\Delta$ axial mass $M_{A\Delta}=1.05\ \mathrm{[GeV]}$. 
The default value of $C_5^A(0)$ is obtained from the Goldberger-Treiman relations \cite{Goldberger:1958tr}:
\begin{eqnarray}
\label{eq:gtrd}
C_5^A(0)=\sqrt{\frac{2}{3}}\frac{f_\pi}{m_\pi}f^\ast\approx 1.2
\end{eqnarray}
which is somewhat higher, than what is used in \cite{Hernandez:2007qq}: $C_5^A(0)\approx 1.15$. 
The authors of \cite{Lalakulich:2006sw} and \cite{Hernandez:2007qq} use $C_5^A(Q^2)=\frac{C_5^A(0)}{(1+Q^2/M_{A\Delta}^2)^2}\frac{1}{1+Q^2/3M_{A\Delta}^2}$.
Because of big uncertainties in axial $N\Delta$ transition, which do not allow to extract any
beyond-dipole behavior, we use the simple dipole form (\ref{eq:c5a}). We include Adler \cite{Adler} relation for $C_4^A$, \textit{i. e.}:
\begin{eqnarray}
\label{eq:c4a}
C_4^A(Q^2)=-\frac{1}{4}C_5^A(Q^2).
\end{eqnarray}
Furthermore, from PCAC hypothesis one can determine:
\begin{eqnarray}
\label{eq:c6a}
C_6^A(Q^2)=\frac{M^2}{m_\pi^2+Q^2}C_5^A(Q^2).
\end{eqnarray}
The $C_3^A$ form factor is considered to be negligibly small, thus we set $C_3^A(Q^2)=0$

\section{Total Cross Section Tables}
\label{sec:tables}
In the following section we present our results in form of the tables.
\begin{landscape}
\begin{table}[htb]
\caption{Total cross-sections in the $^{12}C(\nu_e,e^-)$ scattering in $10^{-38}\ \mathrm{[cm^2]}$.}
\label{nue}\centering 
\begin{tabular}{|c|c|c|c|c|c|c|c|c|c|c|c|c|c|}
\hline
\multirow{3}{*}{$E_\nu$ [GeV]}& \multicolumn{4}{|c|}{6p+6n Free} & \multicolumn{4}{|c|}{Fermi Motion + PB} & \multicolumn{5}{|c|}{Full $\Delta$ In-Medium} \\
&\multicolumn{2}{|c|}{Resonant}&\multicolumn{2}{|c|}{+Background}&\multicolumn{2}{|c|}{Resonant}&\multicolumn{2}{|c|}{+Background}&\multicolumn{2}{|c|}{Resonant}&\multicolumn{2}{|c|}{+Background}& \\
&$\pi^+$&$\pi^0$&$\pi^+$&$\pi^0$&$\pi^+$&$\pi^0$&$\pi^+$&$\pi^0$&$\pi^+$&$\pi^0$&$\pi^+$&$\pi^0$&$\Delta_{pionless}$\\
\hline
0.40 & 0.2274 & 0.0455 & 0.3440 & 0.0888 & 0.3153 & 0.0631 & 0.4183 & 0.0902 & 0.1675 & 0.0335 & 0.2522 & 0.0559 & 0.1412 \\
0.45 & 0.4797 & 0.0959 & 0.6666 & 0.1646 & 0.6063 & 0.1213 & 0.7647 & 0.1646 & 0.3313 & 0.0663 & 0.4664 & 0.1038 & 0.2126 \\
0.50 & 0.8124 & 0.1625 & 1.0674 & 0.2569 & 0.9605 & 0.1921 & 1.1667 & 0.2516 & 0.5430 & 0.1086 & 0.7304 & 0.1619 & 0.2902 \\
0.55 & 1.1980 & 0.2396 & 1.5122 & 0.3581 & 1.3463 & 0.2693 & 1.5886 & 0.3429 & 0.7868 & 0.1574 & 1.0167 & 0.2263 & 0.3682 \\
0.60 & 1.6109 & 0.3222 & 1.9725 & 0.4625 & 1.7449 & 0.3490 & 2.0095 & 0.4353 & 1.0477 & 0.2095 & 1.3132 & 0.2924 & 0.4429 \\
0.65 & 2.0309 & 0.4062 & 2.4282 & 0.5659 & 2.1329 & 0.4266 & 2.4188 & 0.5223 & 1.3137 & 0.2627 & 1.6094 & 0.3560 & 0.5120 \\
0.70 & 2.4433 & 0.4887 & 2.8661 & 0.6657 & 2.5106 & 0.5021 & 2.7844 & 0.6051 & 1.5759 & 0.3152 & 1.8882 & 0.4209 & 0.5745 \\
0.75 & 2.8378 & 0.5676 & 3.2784 & 0.7605 & 2.8616 & 0.5723 & 3.1523 & 0.6821 & 1.8282 & 0.3656 & 2.1566 & 0.4794 & 0.6297 \\
0.80 & 3.2100 & 0.6420 & 3.6634 & 0.8499 & 3.1820 & 0.6364 & 3.4548 & 0.7500 & 2.0667 & 0.4133 & 2.4046 & 0.5338 & 0.6781 \\
0.85 & 3.5571 & 0.7114 & 4.0202 & 0.9338 & 3.4779 & 0.6956 & 3.7404 & 0.8164 & 2.2895 & 0.4579 & 2.6318 & 0.5875 & 0.7199 \\
0.90 & 3.8767 & 0.7753 & 4.3482 & 1.0121 & 3.7432 & 0.7486 & 3.9988 & 0.8747 & 2.4956 & 0.4991 & 2.8464 & 0.6358 & 0.7559 \\
0.95 & 4.1698 & 0.8340 & 4.6506 & 1.0853 & 3.9876 & 0.7975 & 4.2507 & 0.9261 & 2.6859 & 0.5372 & 3.0413 & 0.6764 & 0.7868 \\
1.00 & 4.4389 & 0.8878 & 4.9311 & 1.1543 & 4.2180 & 0.8436 & 4.4820 & 0.9822 & 2.8609 & 0.5722 & 3.2119 & 0.7188 & 0.8131 \\
\hline
\end{tabular}
\end{table}
%\end{landscape}

%\begin{landscape}
\begin{table}[h]
\caption{Total cross-sections in the $^{12}C(\overline{\nu}_e,e^+)$ scattering in $10^{-38}\ \mathrm{[cm^2]}$.}
\label{antinue}\centering 
\begin{tabular}{|c|c|c|c|c|c|c|c|c|c|c|c|c|c|}
\hline
\multirow{3}{*}{$E_\nu$ [GeV]}& \multicolumn{4}{|c|}{6p+6n Free} & \multicolumn{4}{|c|}{Fermi Motion + PB} & \multicolumn{5}{|c|}{Full $\Delta$ In-Medium} \\
&\multicolumn{2}{|c|}{Resonant}&\multicolumn{2}{|c|}{+Background}&\multicolumn{2}{|c|}{Resonant}&\multicolumn{2}{|c|}{+Background}&\multicolumn{2}{|c|}{Resonant}&\multicolumn{2}{|c|}{+Background}& \\
&$\pi^-$&$\pi^0$&$\pi^-$&$\pi^0$&$\pi^-$&$\pi^0$&$\pi^-$&$\pi^0$&$\pi^-$&$\pi^0$&$\pi^-$&$\pi^0$&$\Delta_{pionless}$\\
\hline
0.40 & 0.0475 & 0.0095 & 0.0836 & 0.0305 & 0.0555 & 0.0111 & 0.0848 & 0.0226 & 0.0290 & 0.0058 & 0.0545 & 0.0167 & 0.0216 \\
0.45 & 0.0914 & 0.0183 & 0.1476 & 0.0516 & 0.1023 & 0.0205 & 0.1474 & 0.0396 & 0.0551 & 0.0110 & 0.0960 & 0.0292 & 0.0322 \\
0.50 & 0.1460 & 0.0292 & 0.2245 & 0.0772 & 0.1597 & 0.0319 & 0.2227 & 0.0605 & 0.0891 & 0.0178 & 0.1475 & 0.0452 & 0.0442 \\
0.55 & 0.2088 & 0.0418 & 0.3118 & 0.1065 & 0.2252 & 0.0450 & 0.3067 & 0.0844 & 0.1301 & 0.0260 & 0.2077 & 0.0638 & 0.0573 \\
0.60 & 0.2785 & 0.0557 & 0.4079 & 0.1391 & 0.2992 & 0.0598 & 0.3995 & 0.1110 & 0.1776 & 0.0355 & 0.2763 & 0.0849 & 0.0713 \\
0.65 & 0.3543 & 0.0709 & 0.5120 & 0.1745 & 0.3759 & 0.0752 & 0.4997 & 0.1398 & 0.2309 & 0.0462 & 0.3528 & 0.1078 & 0.0859 \\
0.70 & 0.4351 & 0.0870 & 0.6231 & 0.2124 & 0.4630 & 0.0926 & 0.6075 & 0.1702 & 0.2893 & 0.0579 & 0.4350 & 0.1329 & 0.1010 \\
0.75 & 0.5204 & 0.1041 & 0.7408 & 0.2525 & 0.5515 & 0.1103 & 0.7199 & 0.2011 & 0.3522 & 0.0704 & 0.5246 & 0.1598 & 0.1163 \\
0.80 & 0.6099 & 0.1220 & 0.8649 & 0.2945 & 0.6416 & 0.1283 & 0.8391 & 0.2354 & 0.4191 & 0.0838 & 0.6192 & 0.1879 & 0.1316 \\
0.85 & 0.7026 & 0.1405 & 0.9945 & 0.3382 & 0.7392 & 0.1478 & 0.9624 & 0.2689 & 0.4890 & 0.0978 & 0.7205 & 0.2164 & 0.1470 \\
0.90 & 0.7980 & 0.1596 & 1.1295 & 0.3835 & 0.8359 & 0.1672 & 1.0959 & 0.3064 & 0.5613 & 0.1123 & 0.8286 & 0.2475 & 0.1623 \\
0.95 & 0.8956 & 0.1791 & 1.2695 & 0.4301 & 0.9342 & 0.1868 & 1.2316 & 0.3407 & 0.6358 & 0.1272 & 0.9393 & 0.2788 & 0.1775 \\
1.00 & 0.9951 & 0.1990 & 1.4145 & 0.4780 & 1.0378 & 0.2076 & 1.3656 & 0.3795 & 0.7121 & 0.1424 & 1.0583 & 0.3108 & 0.1924 \\
\hline
\end{tabular}
\end{table}

\begin{table}[h]
\caption{Total cross-sections in the $^{12}C(\nu_\mu,\mu^-)$ scattering in $10^{-38}\ \mathrm{[cm^2]}$.}
\label{numu}\centering 
\begin{tabular}{|c|c|c|c|c|c|c|c|c|c|c|c|c|c|}
\hline
\multirow{3}{*}{$E_\nu$ [GeV]}& \multicolumn{4}{|c|}{6p+6n Free} & \multicolumn{4}{|c|}{Fermi Motion + PB} & \multicolumn{5}{|c|}{Full $\Delta$ In-Medium} \\
&\multicolumn{2}{|c|}{Resonant}&\multicolumn{2}{|c|}{+Background}&\multicolumn{2}{|c|}{Resonant}&\multicolumn{2}{|c|}{+Background}&\multicolumn{2}{|c|}{Resonant}&\multicolumn{2}{|c|}{+Background}& \\
&$\pi^+$&$\pi^0$&$\pi^+$&$\pi^0$&$\pi^+$&$\pi^0$&$\pi^+$&$\pi^0$&$\pi^+$&$\pi^0$&$\pi^+$&$\pi^0$&$\Delta_{pionless}$\\
\hline
0.40 & 0.0276 & 0.0055 & 0.0635 & 0.0202 & 0.0677 & 0.0135 & 0.1102 & 0.0228 & 0.0388 & 0.0078 & 0.0717 & 0.0153 & 0.0721 \\
0.45 & 0.1378 & 0.0276 & 0.2406 & 0.0658 & 0.2528 & 0.0506 & 0.3538 & 0.0749 & 0.1316 & 0.0263 & 0.2107 & 0.0458 & 0.1317 \\
0.50 & 0.4078 & 0.0816 & 0.6007 & 0.1507 & 0.5576 & 0.1115 & 0.7228 & 0.1545 & 0.2981 & 0.0596 & 0.4348 & 0.0956 & 0.2063 \\
0.55 & 0.7792 & 0.1558 & 1.0495 & 0.2531 & 0.9331 & 0.1866 & 1.1481 & 0.2480 & 0.5197 & 0.1039 & 0.7122 & 0.1573 & 0.2869 \\
0.60 & 1.1890 & 0.2378 & 1.5196 & 0.3594 & 1.3370 & 0.2674 & 1.5934 & 0.3425 & 0.7725 & 0.1545 & 1.0116 & 0.2244 & 0.3667 \\
0.65 & 1.6128 & 0.3226 & 1.9898 & 0.4655 & 1.7418 & 0.3484 & 2.0115 & 0.4358 & 1.0382 & 0.2076 & 1.3167 & 0.2904 & 0.4417 \\
0.70 & 2.0407 & 0.4081 & 2.4533 & 0.5704 & 2.1309 & 0.4262 & 2.4195 & 0.5230 & 1.3044 & 0.2609 & 1.6114 & 0.3554 & 0.5099 \\
0.75 & 2.4628 & 0.4926 & 2.8998 & 0.6718 & 2.4985 & 0.4997 & 2.7860 & 0.6017 & 1.5630 & 0.3126 & 1.8779 & 0.4188 & 0.5705 \\
0.80 & 2.8596 & 0.5719 & 3.3105 & 0.7661 & 2.8390 & 0.5678 & 3.1156 & 0.6740 & 1.8092 & 0.3618 & 2.1372 & 0.4754 & 0.6238 \\
0.85 & 3.2263 & 0.6453 & 3.6855 & 0.8533 & 3.1519 & 0.6304 & 3.4359 & 0.7380 & 2.0404 & 0.4081 & 2.3644 & 0.5294 & 0.6700 \\
0.90 & 3.5638 & 0.7128 & 4.0288 & 0.9345 & 3.4382 & 0.6876 & 3.6895 & 0.8048 & 2.2549 & 0.4510 & 2.5930 & 0.5776 & 0.7098 \\
0.95 & 3.8729 & 0.7746 & 4.3430 & 1.0100 & 3.6954 & 0.7391 & 3.9564 & 0.8624 & 2.4532 & 0.4906 & 2.7903 & 0.6250 & 0.7439 \\
1.00 & 4.1550 & 0.8310 & 4.6313 & 1.0805 & 3.9245 & 0.7849 & 4.1873 & 0.9113 & 2.6352 & 0.5270 & 2.9678 & 0.6661 & 0.7730 \\
\hline
\end{tabular}
\end{table}

\begin{table}[h]
\caption{Total cross-sections in the $^{12}C(\overline{\nu}_\mu,\mu^+)$ scattering in $10^{-38}\ \mathrm{[cm^2]}$.}
\label{antinumu}\centering 
\begin{tabular}{|c|c|c|c|c|c|c|c|c|c|c|c|c|c|}
\hline
\multirow{3}{*}{$E_\nu$ [GeV]}& \multicolumn{4}{|c|}{6p+6n Free} & \multicolumn{4}{|c|}{Fermi Motion + PB} & \multicolumn{5}{|c|}{Full $\Delta$ In-Medium} \\
&\multicolumn{2}{|c|}{Resonant}&\multicolumn{2}{|c|}{+Background}&\multicolumn{2}{|c|}{Resonant}&\multicolumn{2}{|c|}{+Background}&\multicolumn{2}{|c|}{Resonant}&\multicolumn{2}{|c|}{+Background}& \\
&$\pi^-$&$\pi^0$&$\pi^-$&$\pi^0$&$\pi^-$&$\pi^0$&$\pi^-$&$\pi^0$&$\pi^-$&$\pi^0$&$\pi^-$&$\pi^0$&$\Delta_{pionless}$\\
\hline
0.40 & 0.0049 & 0.0010 & 0.0173 & 0.0096 & 0.0076 & 0.0015 & 0.0186 & 0.0059 & 0.0045 & 0.0009 & 0.0141 & 0.0052 & 0.0079 \\
0.45 & 0.0212 & 0.0042 & 0.0495 & 0.0223 & 0.0291 & 0.0058 & 0.0534 & 0.0159 & 0.0150 & 0.0030 & 0.0361 & 0.0127 & 0.0146 \\
0.50 & 0.0587 & 0.0117 & 0.1082 & 0.0423 & 0.0673 & 0.0135 & 0.1084 & 0.0315 & 0.0351 & 0.0070 & 0.0716 & 0.0240 & 0.0240 \\
0.55 & 0.1089 & 0.0218 & 0.1812 & 0.0671 & 0.1193 & 0.0239 & 0.1794 & 0.0514 & 0.0645 & 0.0129 & 0.1189 & 0.0390 & 0.0354 \\
0.60 & 0.1649 & 0.0330 & 0.2623 & 0.0952 & 0.1817 & 0.0363 & 0.2614 & 0.0748 & 0.1020 & 0.0204 & 0.1765 & 0.0568 & 0.0483 \\
0.65 & 0.2286 & 0.0457 & 0.3545 & 0.1271 & 0.2530 & 0.0506 & 0.3546 & 0.1009 & 0.1469 & 0.0294 & 0.2432 & 0.0774 & 0.0624 \\
0.70 & 0.3058 & 0.0612 & 0.4643 & 0.1638 & 0.3310 & 0.0662 & 0.4531 & 0.1286 & 0.1981 & 0.0396 & 0.3167 & 0.1000 & 0.0773 \\
0.75 & 0.3967 & 0.0793 & 0.5887 & 0.2042 & 0.4151 & 0.0830 & 0.5619 & 0.1597 & 0.2550 & 0.0510 & 0.3996 & 0.1244 & 0.0927 \\
0.80 & 0.4903 & 0.0981 & 0.7149 & 0.2458 & 0.5036 & 0.1007 & 0.6748 & 0.1913 & 0.3166 & 0.0633 & 0.4871 & 0.1509 & 0.1083 \\
0.85 & 0.5841 & 0.1168 & 0.8424 & 0.2883 & 0.5962 & 0.1192 & 0.7937 & 0.2237 & 0.3823 & 0.0765 & 0.5811 & 0.1783 & 0.1241 \\
0.90 & 0.6792 & 0.1358 & 0.9733 & 0.3321 & 0.6922 & 0.1384 & 0.9189 & 0.2578 & 0.4512 & 0.0902 & 0.6796 & 0.2079 & 0.1398 \\
0.95 & 0.7763 & 0.1553 & 1.1084 & 0.3771 & 0.7901 & 0.1580 & 1.0436 & 0.2936 & 0.5229 & 0.1046 & 0.7845 & 0.2372 & 0.1555 \\
1.00 & 0.8748 & 0.1750 & 1.2473 & 0.4232 & 0.8896 & 0.1779 & 1.1769 & 0.3287 & 0.5965 & 0.1193 & 0.8926 & 0.2685 & 0.1709 \\
\hline
\end{tabular}
\end{table}
\end{landscape}

\bibliography{bibliography}{}

\begin{thebibliography}{10}

\bibitem{Nieves:2011pp}
J.~Nieves, I.~Ruiz~Simo, and M.~J. Vicente~Vacas.
\newblock Phys.\ Rev.\ C, {\bf 83}:045501, 2011\relax
\relax
\bibitem{AguilarArevalo:2010zc}
A.~A.~Aguilar-Arevalo {\it et al.}~[MiniBooNE~Collaboration].
\newblock Phys.\ Rev.\ D, {\bf 81}:092005, 2010\relax
\relax
\bibitem{problems_RS}
K.M. Graczyk and J.T. Sobczyk.
\newblock Phys. Rev. D, {\bf 77}:053001, 2008.
\newblock erratum-ibid. D{\bf 79} (2009) 079903\relax
\relax
\bibitem{en_rec}
M.~Martini, M.~Ericson, and G.~Chanfray.
\newblock Phys.\ Rev.\ D, {\bf 85}:093012, 2012.
\newblock ;\\J. Nieves, F. Sanchez, I. Ruiz Simo, and M. J. Vicente Vacas.
  Phys. Rev. D, {\bf 85}:113008, 2012 ;\\O. Lalakulich and U. Mosel.
  arXiv:1208.3678 [nucl-th]\relax
\relax
\bibitem{Tiator:2011pw}
L.~Tiator, D.~Drechsel, S.~S. Kamalov, and M.~Vanderhaeghen.
\newblock Eur.\ Phys.\ J.\ ST, {\bf 198}:141, 2011\relax
\relax
\bibitem{Drechsel:2007if}
D.~Drechsel, S.~S. Kamalov, and L.~Tiator.
\newblock Eur.\ Phys.\ J.\ A, {\bf 34}:69, 2007\relax
\relax
\bibitem{Rodriguez:2008aa}
A.~Rodriguez and L~{\it et al.} [K2K~Collaboration] Whitehead.
\newblock Phys.\ Rev.\ D, {\bf 78}:032003, 2008\relax
\relax
\bibitem{Mariani:2010ez}
C.~Mariani {\it et al.}~[K2K~Collaboration].
\newblock Phys.\ Rev.\ D, {\bf 83}:054023, 2011\relax
\relax
\bibitem{AguilarArevalo:2010bm}
A.~A.~Aguilar-Arevalo {\it et al.}~[MiniBooNE~Collaboration].
\newblock Phys.\ Rev.\ D, {\bf 83}:052007, 2011\relax
\relax
\bibitem{AguilarArevalo:2010xt}
A.~A.~Aguilar-Arevalo {\it et al.}~[MiniBooNE~Collaboration].
\newblock Phys.\ Rev.\ D, {\bf 83}:052009, 2011\relax
\relax
\bibitem{Barish:1978pj}
S.~J. Barish, M.~Derrick, T.~Dombeck, L.~G. Hyman, K.~Jaeger, B.~Musgrave,
  P.~Schreiner, and R.~{\it et al.} Singer.
\newblock Phys.\ Rev.\ D, {\bf 19}:2521, 1979\relax
\relax
\bibitem{Radecky:1981fn}
G.~M. Radecky, V.~E. Barnes, D.~D. Carmony, A.~F. Garfinkel, M.~Derrick,
  E.~Fernandez, L.~Hyman, and G.~{\it et al.} Levman.
\newblock Phys.\ Rev.\ D, {\bf 26}:3297, 1982.
\newblock [Erratum-ibid.\ D {\bf 26} (1982) 3297]\relax
\relax
\bibitem{Kitagaki:1990vs}
T.~Kitagaki, H.~Yuta, S.~Tanaka, A.~Yamaguchi, K.~Abe, K.~Hasegawa, K.~Tamai,
  and H.~{\it et al.} Sagawa.
\newblock Phys.\ Rev.\ D, {\bf 42}:1331, 1990\relax
\relax
\bibitem{Hernandez:2007qq}
E.~Hernandez, J.~Nieves, and M.~Valverde.
\newblock Phys.\ Rev.\ D, {\bf 76}:033005, 2007\relax
\relax
\bibitem{Lalakulich:2010ss}
O.~Lalakulich, T.~Leitner, O.~Buss, and U.~Mosel.
\newblock Phys.\ Rev.\ D, {\bf 82}:093001, 2010\relax
\relax
\bibitem{Oset:1987re}
E.~Oset and L.~L. Salcedo.
\newblock Nucl.\ Phys.\ A, {\bf 468}:631, 1987\relax
\relax
\bibitem{Singhy}
S.~K. Singh, M.~J. Vicente-Vacas, and E.~Oset.
\newblock Phys.\ Lett.\ B, {\bf 416}:23, 1998.
\newblock [Erratum-ibid.\ B {\bf 423} (1998) 428]; \\S. Ahmad, M. Sajjad Athar
  and S. K. Singh. Phys. Rev. D, {\bf 74}:073008, 2006; \\M. Sajjad Athar, S.
  Ahmad, and S. K. Singh. Nucl. Phys. A, {\bf 782}:179, 2007; \\S. K. Singh, M.
  Sajjad Athar, and S. Ahmed. AIP Conf. Proc., {\bf 967}:182, 2007; \\M. Sajjad
  Athar, S. Chauhan, and S. K. Singh. J. Phys. G, {\bf 37}:015005, 2010; \\M.
  Sajjad Athar, S. Chauhan, and S. K. Singh. Eur. Phys. J. A, {\bf 43}:209,
  2010\relax
\relax
\bibitem{Beringer:1900zz}
J.~Beringer {\it et al.}~[Particle Data Group~Collaboration].
\newblock Phys.\ Rev.\ D, {\bf 86}:010001, 2012\relax
\relax
\bibitem{Gil:1997bm}
A.~Gil, J.~Nieves, and E.~Oset.
\newblock Nucl.\ Phys.\ A, {\bf 627}:543, 1997\relax
\relax
\bibitem{GSL}
M.~Galassi {\it et al.}
\newblock {\em GNU Scientific Library Reference Manual (3rd Ed.)}.
\newblock Network Theory Ltd., 2009\relax
\relax
\bibitem{Antonello:2009ca}
M.~Antonello, V.~Caracciolo, G.~Christodoulou, J.~Dobson, E.~Frank, T.~Golan,
  V.~Lee, and S.~{\it et al.} Mania.
\newblock Acta Phys.\ Polon.\ B, {\bf 40}:2519, 2009\relax
\relax
\bibitem{mosel}
O.~Lalakulich, K.~Gallmeister, T.~Leitner, and U.~Mosel.
\newblock AIP Conf.Proc., {\bf 1405}:127, 2011\relax
\relax
\bibitem{mec}
J.~Marteau, J.~Delorme, and M.~Ericson.
\newblock Nucl.\ Instrum.\ Meth.\ A, {\bf 451}:76, 2000.
\newblock ;\\ M. Martini, M. Ericson, G. Chanfray, and J. Marteau. Phys. Rev.
  C, {\bf 80}:065501, 2009;\\ J. Nieves, I. Ruiz Simo, and M. J. Vicente Vacas.
  Phys. Lett. B, {\bf 707}:72, 2012\relax
\relax
\bibitem{nuwro}
C.~Juszczak, J.~A. Nowak, and J.~T. Sobczyk.
\newblock Nucl.\ Phys.\ Proc.\ Suppl., {\bf 159}:211, 2006.
\newblock \\J. A. Nowak. Phys. Scripta T, {\bf 127}:70, 2006;\\C. Juszczak.
  Acta Phys. Polon. B, {\bf 40}:2507, 2009\relax
\relax
\bibitem{Manley:1992yb}
D.~M. Manley and E.~M. Saleski.
\newblock Phys.\ Rev.\ D, {\bf 45}:4002, 1992\relax
\relax
\bibitem{Leitner:2009phd}
T.~Leitner.
\newblock {\em Neutrino-Nucleus Interactions in a Coupled-Channel Hadronic
  Transport Model}.
\newblock PhD thesis, Justus-Liebig-Universitaet Giessen, Fachbereich 07
  (Mathematik und Informatik, Physik, Geographie), Institut fuer Theoretische
  Physik, 2007\relax
\relax
\bibitem{DeJager:1987qc}
H.~De~Vries, C.~W. De~Jager, and C.~De~Vries.
\newblock Atom.\ Data Nucl.\ Data Tabl., {\bf 36}:495, 1987\relax
\relax
\bibitem{GarciaRecio:1991wk}
C.~Garcia-Recio, J.~Nieves, and E.~Oset.
\newblock Nucl.\ Phys.\ A, {\bf 547}:473, 1992\relax
\relax
\bibitem{Galster:1971kv}
S.~Galster, H.~Klein, J.~Moritz, K.~H. Schmidt, D.~Wegener, and J.~Bleckwenn.
\newblock Nucl.\ Phys.\ B, {\bf 32}:221, 1971\relax
\relax
\bibitem{Lalakulich:2006sw}
O.~Lalakulich, E.~A. Paschos, and G.~Piranishvili.
\newblock Phys.\ Rev.\ D, page 014009, 2006\relax
\relax
\bibitem{Goldberger:1958tr}
M.~L. Goldberger and S.~B. Treiman.
\newblock Phys.\ Rev., {\bf 110}:1178, 1958\relax
\relax
\bibitem{Adler}
S.~L. Adler.
\newblock Annals Phys., {\bf 50}:189, 1968\relax
\relax
\end{thebibliography}
\bibliographystyle{unsort}

\end{document}